\documentclass[aps,prb,reprint,superscriptaddress,footinbib,longbibliography]{revtex4-2}
\usepackage{graphicx}
\usepackage{dcolumn}
\usepackage{bm}
\usepackage{amsmath}
\usepackage{amssymb}
\usepackage{amsthm}
\usepackage{braket}
\usepackage{physics}
\usepackage{xcolor}
\usepackage{hyperref}
\usepackage{array}
\usepackage{booktabs}
\usepackage[english]{babel}
\usepackage{mathtools}

\hypersetup{
    colorlinks,
    linkcolor={blue!70!black},
    citecolor={blue!70!black},
    urlcolor ={blue!70!black}
}

\newcommand{\ii}{{\mathrm{i}}}

\newcommand{\dk}{\Delta_k}

\newcommand{\Ae}{{\mathcal{A}^{e}}}
\newcommand{\Ah}{{\mathcal{A}^{h}}}
\newcommand{\Dmat}{{\mathcal{D}}}
\newcommand{\Abar}{{\overline{\mathcal{A}}}}
\newcommand{\BMx}{{B_{\mathcal{M}_x}}}

\newcommand{\Me}{{M_e}}
\newcommand{\Mh}{{M_h}}
\newcommand{\Mrelmat}{{M_{\mathrm{rel}}}}

\newcommand{\GS}{{\ket{\mathrm{GS}}}}

\newcommand{\Xe}{{X_e}}
\newcommand{\Xh}{{X_h}}

\newcommand{\BareXe}{{\mathcal{X}_e}}
\newcommand{\BareXh}{{\mathcal{X}_h}}

\newcommand{\tXe}{{\widetilde{X}_e}}
\newcommand{\tXh}{{\widetilde{X}_h}}

\newcommand{\tBareXe}{{\widetilde{\mathcal{X}}_e}}
\newcommand{\tBareXh}{{\widetilde{\mathcal{X}}_h}}

\newcommand{\tXeC}[1]{{\widetilde{X}^{e}_{C}(#1)}}
\newcommand{\tXeS}[1]{{\widetilde{X}^{e}_{S}(#1)}}
\newcommand{\tXhC}[1]{{\widetilde{X}^{h}_{C}(#1)}}
\newcommand{\tXhS}[1]{{\widetilde{X}^{h}_{S}(#1)}}

\newcommand{\Lexc}{{L_{\mathrm{exc}}}}
\newcommand{\LIF}{{\mu}}

\newcommand{\Pexc}{{P_{\mathrm{exc}}}}

\newcommand{\Gam}[1]{{\Gamma_{#1}}}

\newcommand{\Lsize}{{L}}

\newcommand{\psch}{p}
\newcommand{\esch}{\xi}
\newcommand{\satresidual}{\mathcal{S}}

\newcommand{\trop}{\tilde{r}}
\newcommand{\tRop}{\widetilde{R}}

\begin{document}

\onecolumngrid
\preprint{APS/123-QED}

\title{A Spatial Localizer for Constituent-Resolved Exciton Wannier Functions}

\author{Haylen Gerhard}
\affiliation{Department of Physics, Emory University, Atlanta, Georgia 30322, USA}

\author{Wladimir A. Benalcazar}
\affiliation{Department of Physics, Emory University, Atlanta, Georgia 30322, USA}

\begin{abstract}
Excitons are composite quasiparticles: beyond a center-of-mass position, each carries an internal electron-hole dipole governing its coupling to electric fields and other excitons. Exciton Wannier functions locally represent exciton bands, but resolving this dipole requires localizing the electron and hole simultaneously. We show that, in one dimension, the projected electron and hole position operators fail to commute when the covariant derivative of the quantum geometric dipole (QGD) matrix (the difference between the hole and electron non-Abelian Berry connections) is nonzero. This precludes a common eigenbasis and bounds the joint electron-hole spread from below. For one band, the internal dipole is gauge invariant and center-of-mass methods suffice; for multiple bands, no existing construction yields a gauge minimizing both position uncertainties. We introduce an ``exciton spatial localizer,'' a Hermitian operator embedding both projected positions in a Clifford-algebra structure. Its spectral minima locate the exciton's center-of-mass and dipole coordinates, while its eigenvectors yield exciton Wannier functions jointly localized in electron and hole coordinates without gauge fixing, an ansatz, or iterative optimization. In an interacting bilayer model, combined reflection--time-reversal symmetry or a nonsymmorphic particle--hole symmetry forces the QGD matrix to be traceless at every momentum while allowing it to remain nonzero. A two-band exciton subspace with zero net internal dipole then decomposes into a symmetry-related pair of exciton Wannier functions with opposite center-of-mass positions and internal dipoles. Adding the interlayer dipole as a Clifford component further separates intralayer and interlayer exciton Wannier functions in a six-band subspace.
\end{abstract}

\maketitle

\newpage

\twocolumngrid
\section{Introduction}
\label{sec:introduction}
An exciton is a bound state of an electron and a hole. The modern theory of polarization and band geometry provides a useful framework for understanding the geometric and topological properties of excitons~\cite{yao2008,srivastava2015,zhou2015,kwan2021,haber2023,davenport2024,davenport2026berryology}.
Exciton bands are Bloch bands of these composite quasiparticles labeled by the total exciton momentum $Q$. Their dispersion $E(Q)$ governs center of mass (COM) transport, and the COM position space coordinate $R$ describes the exciton's location in moir\'e or effective trapping potentials~\cite{yao2008,wu2017,kwan2021}. In many experiments, however, excitons are distinguished not only by where the bound excitation is located, but also by how the electron and hole are arranged within it. This internal electron-hole structure determines permanent dipole moments and can govern Stark shifts, optical response, radiative lifetimes, transport, and dipole-mediated exciton--exciton interactions in interlayer, hybrid, and moir\'e exciton systems~\cite{barre2022opticalabsorption,schwandt2025ferroelectric,tagarelli2023hybrid,jiang2021interlayer,wu2018opticalabsorption,gotting2022moire}.

A real-space description of excitons would ideally be \emph{constituent-resolved}: for each localized exciton, it would track both the position of its electron, $x_e$, and that of its hole, $x_h$, or, equivalently, its COM coordinate $R=(x_e+x_h)/2$ and its internal dipole $r=x_h-x_e$ (expressed in $\abs{e}=1$ units). In the search for such a description, two complementary constructions of exciton Wannier functions (eWFs) have been obtained. The first extends the maximally localized Wannier function framework~\cite{marzari1997,pizzi2020} to a multiband exciton subspace by choosing a smooth $Q$-dependent gauge that minimizes the spread of a selected weighted electron-hole coordinate, including the equal-weight COM coordinate~\cite{haber2023}; we refer to these states as COM-maximally-localized.
The second uses constituent projected-position operators and Wilson loops to construct electron- and hole-localized eWFs for an isolated exciton band~\cite{davenport2026berryology}. The former accommodates composite, overlapping, and degenerate exciton bands but localizes only one chosen weighted coordinate, whereas the latter resolves the constituent positions separately but does not provide a common multiband gauge that jointly localizes the electron and the hole. Throughout this work, we focus on excitons in one dimension (1D).

The underlying obstruction to a constituent-resolved description is that, after projection onto the exciton subspace, the electron and hole position operators, $\tBareXe$ and $\tBareXh$, need not commute. Equivalently, the projected COM and internal-dipole operators need not commute, since $[\tBareXe,\tBareXh]=[\tRop,\trop]$. As we show below, this commutator is controlled by the \emph{quantum geometric dipole} (QGD) matrix $\Dmat(Q)$, defined as the difference between the hole- and electron-resolved non-Abelian Berry connections: the projected positions fail to commute whenever the covariant derivative of $\Dmat(Q)$ is nonzero. A nonzero commutator precludes a common eigenbasis of the two position operators and bounds their joint spread from below. The problem then becomes considerably more complicated: rather than a simultaneous diagonalization, it amounts to choosing a gauge for the exciton bands that minimizes the uncertainties of both position operators simultaneously. This situation resembles that of energy bands in insulators in 2D or beyond, where the projected position operators along different directions generally do not commute~\cite{marzari1997}. 

How consequential the obstruction is depends on the number of exciton bands. For a single isolated exciton band, the QGD is a gauge-invariant scalar, so both the internal dipole $r^W$ of the eWF and its spread are fixed; only the COM spread can be minimized by the choice of gauge, and finding the COM coordinate $R^W$ suffices. This is achieved by current methods in 1D~\cite{haber2023, davenport2026berryology}. For multiple exciton bands, in contrast, $\Dmat(Q)$ transforms covariantly under band mixing, and finding the collection of Wannier coordinates $\{R^W,r^W\}$ requires a gauge that balances the localization of both sets of coordinates at once. No existing method provides such a gauge, and Wilson loop methods become cumbersome in this setting and have not hitherto been employed.

The multiband case is also where constituent resolution has qualitatively new consequences. Mirror symmetry $\mathcal{M}_x$ provides a minimal example. For an isolated exciton band, a mirror-symmetric eWF is mapped to itself under $\mathcal{M}_x$ up to a lattice translation; thus, its COM lies at a mirror-invariant coordinate and its internal dipole vanishes~\cite{davenport2026berryology}. With multiple exciton bands, however, $\mathcal{M}_x$ may act on the entire Wannier configuration rather than on each eWF separately. This allows a mirror-symmetric exciton subspace with zero net internal dipole whose individual eWFs nevertheless carry nonzero internal dipoles. This structure is invisible to the gauge-invariant trace of the QGD matrix; determining how the vanishing net internal dipole is distributed among individual eWFs requires a localization framework that resolves both constituents simultaneously in each eWF.

In this work, we present a method for finding maximally localized eWFs, i.e., those that minimize the total joint electron-hole spread within the exciton subspace, and their accompanying Wannier coordinates, $(R^W,r^W)$. Our method is based on the construction of an ``exciton spatial localizer,'' a Hermitian operator built from the projected electron and hole position operators embedded in a Clifford algebra, as recently conceived for the 2D insulator problem~\cite{gerhard2026}, and which itself draws inspiration from earlier work on the problem of simultaneous diagonalization of non-commuting operators~\cite{loring2015,cerjan2022}. The exciton spatial localizer reframes the localization problem so that a gauge-fixing procedure is not necessary: it recasts the search for an optimal gauge as an eigenvalue problem parametrized by trial electron and hole positions. Diagonalizing the localizer yields candidate Wannier coordinates and their associated localized states without the need for an ansatz, gauge fixing, or iterative gauge optimization; a subsequent L\"owdin orthogonalization of these states yields the orthonormal eWFs. A key distinction between the 2D insulator spatial localizer and the 1D exciton spatial localizer is that, in the latter, the internal dipole coordinate $r^W$ is an internal degree of freedom not associated with a Berry connection but, instead, with the QGD.

Our main results are as follows. First, we show that the commutator of the projected electron and hole positions equals the covariant derivative of the QGD matrix, which identifies when joint localization is obstructed and bounds the joint spread from below. Second, we construct the exciton spatial localizer and use it to obtain constituent-resolved eWFs for single and multiple exciton bands with nontrivial COM coordinates and internal dipoles. Third, in an interacting bilayer Su--Schrieffer--Heeger (SSH) model, we show that $\mathcal{M}_x\mathcal{T}$, with $\mathcal{T}$ the spinless time-reversal operator, or a nonsymmorphic particle-hole symmetry forces the QGD matrix to be traceless at every $Q$ without forcing it to vanish, so that a two-band exciton subspace with zero net internal dipole decomposes into a mirror-related pair of eWFs with equal and opposite internal dipoles. Fourth, we present an extended version of the exciton spatial localizer that additionally resolves the interlayer component of the internal dipole, a distinction relevant to separating interlayer from in-plane dipoles in bilayer material platforms, including hexagonal BN-encapsulated $\text{WSe}_2$ homobilayers and other transition metal dichalcogenide heterostructures~\cite{barre2022opticalabsorption,jiang2021interlayer,tagarelli2023hybrid,schwandt2025ferroelectric}. Our construction assumes an energetically isolated set of exciton bands; at finite system size, the candidate coordinates obtained from the localizer agree with the Wannier coordinates evaluated from the eWFs up to small differences that vanish in the thermodynamic limit.

The paper is organized as follows. In Sec.~\ref{sec:multiband-exciton-geometry}, we introduce the electron- and hole-resolved non-Abelian Berry connections and the QGD matrix, and define the exciton spatial localizer, using a symmetry-broken isolated exciton band as an example. In Sec.~\ref{sec:multiband_excitons}, we determine the symmetry constraints on multiband exciton subspaces and apply the localizer, and its layer-resolved extension, to two- and six-band subspaces of the bilayer SSH model. Section~\ref{sec:discussion-conclusions} presents a discussion and outlook.

\section{Constituent-Resolved Exciton Localization}
\label{sec:multiband-exciton-geometry}

In this section we develop the machinery used to resolve both constituents of a localized exciton. We first specify the active exciton subspace and its constituent coordinates, then introduce the projected electron and hole position operators and their matrix-valued Berry connections. We finally formulate the spatial localizer as a pathway to joint localization.

\subsection{Exciton subspace and constituent coordinates}
\label{subsec:active-exciton-bands}

We work in the single-exciton Hilbert space generated by particle-hole states
\begin{equation}\nonumber
\ket{Q,k}
=
c^\dagger_{k+Q}v_k\GS,
\label{eq:single-exciton-basis}
\end{equation}
where $c$ and $v$ respectively annihilate conduction and valence band electrons, $Q$ is the many-body exciton momentum, $k$ labels the internal electron-hole configuration, and $\GS$ is the reference state with the active valence band filled and active conduction band empty. In this convention, the conduction electron carries momentum $k+Q$, while the valence hole carries momentum $-k$, so that $Q$ is the total exciton momentum.

\begin{figure}[!htb]
\centering
\includegraphics[width=1\linewidth]{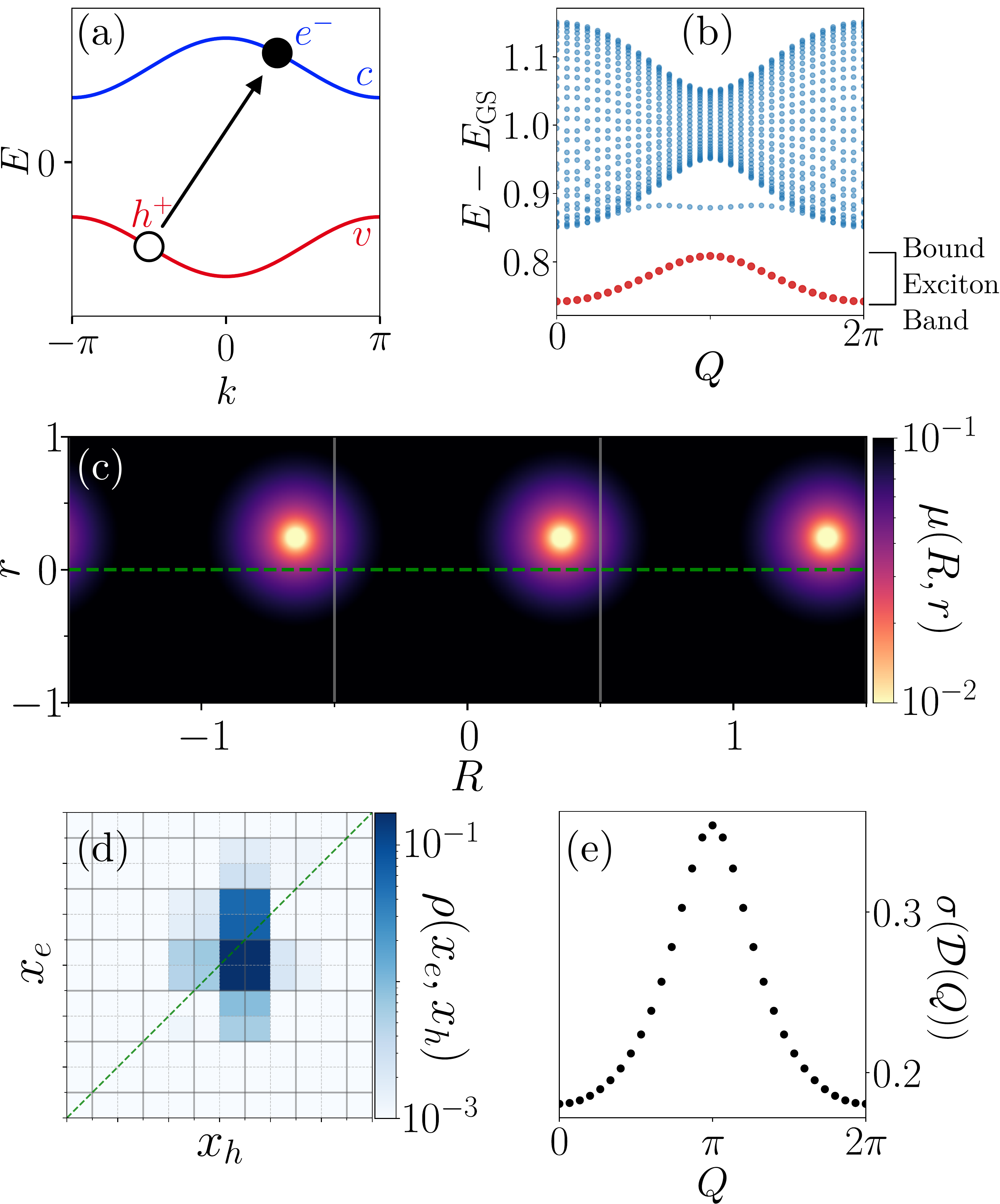}
\caption{{\bf Constituent-resolved localization of a symmetry-broken exciton band.} (a) Schematic of exciton formation by promoting a valence-band electron into a conduction band, leaving behind a valence hole. (b) Excitation spectrum in the single-exciton Hilbert space, where the isolated active exciton band is highlighted in red and $E_{\mathrm{GS}}$ is the energy of the reference state $\GS$. (c) LIF $\mu(R,r)$ with a single minimum per unit cell. The vertical gray lines denote unit cell boundaries. (d) Joint density $\rho(x_e,x_h)$ of the eWF constructed using the spatial localizer. The solid gray lines denote unit cell boundaries and the dashed gray lines denote orbital boundaries within the unit cell. The green dashed lines in (c,d) denote zero internal dipole, corresponding to $r=0$ ($x_e=x_h$). (e) $\Dmat(Q)$ values of the active exciton band, with the nonzero dispersion in $Q$ indicating $\comm{\tBareXe}{\tBareXh}\neq0$.}
\label{fig:figure1}
\end{figure}

Within this sector, the valence hole is a missing electron in the otherwise-filled valence band. Therefore, the repulsive density-density interaction enters the electron-hole channel with opposite sign, producing a net attraction that can bind the two constituents. Figure~\ref{fig:figure1}(a) schematically illustrates the resulting promotion of a valence electron into a conduction band, while Fig.~\ref{fig:figure1}(b) shows a corresponding isolated bound exciton band separated from the particle-hole continuum. The spectrum in Fig.~\ref{fig:figure1}(b) arises from a symmetry-broken parameterization of a bilayer Su–Schrieffer–Heeger (SSH) model (see Sec.~\ref{sec:multiband_excitons} and Appendix~\ref{app:model} for model details). 

More generally, we focus on an isolated group of $N_{\mathrm{exc}}$ exciton bands, which we hereon refer to as the active exciton bands. At each $Q$, the corresponding Bloch exciton states span an $N_{\mathrm{exc}}$-dimensional active subspace and can be written as

\begin{equation}\nonumber
    \ket{\Psi_{\alpha,Q}}
    =
    \sum_k
    \phi^Q_{\alpha,k}
    \ket{Q,k},
    \label{eq:active-exciton-bloch-state}
\end{equation}
where $\alpha=1,\dots,N_{\mathrm{exc}}$ labels the active exciton bands and the coefficients $\phi^Q_{\alpha,k}$ satisfy the orthonormality condition $\sum_k \bar{\phi}^Q_{\alpha,k} \phi^Q_{\beta,k} = \delta_{\alpha\beta}$.

The choice of active exciton frame is not unique. At each $Q$, the states may be rotated by a smooth unitary matrix within this subspace,
\begin{equation}\nonumber
    \ket{\Psi_{\alpha,Q}}
    \rightarrow
    \sum_{\beta}
    \ket{\Psi_{\beta,Q}}
    U_{\beta\alpha}(Q),
    \label{eq:exciton-gauge-freedom}
\end{equation}
which is directly analogous to the gauge freedom of isolated groups of single-particle bands in ordinary Wannier theory~\cite{marzari1997}. eWFs are constructed from smooth, periodic choices of this active exciton frame over $Q$. Such $Q$-dependent unitary mixing has been implemented explicitly for composite sets of overlapping and degenerate exciton bands in the construction of COM-maximally-localized eWFs~\cite{haber2023}. For notational clarity, the analytic development here retains one single-particle parent conduction band and one single-particle parent valence band from which excitons are considered. With multiple parent bands, $k$ is supplemented by conduction- and valence-band indices. 

Because an exciton is a composite quasiparticle, a single eWF $\ket{W_\nu}$, for $\nu=1,...,N_{\mathrm{exc}}$, can be assigned two constituent centers, $(x_{e,\nu}^W,x_{h,\nu}^W)$, defined as the first moments of the corresponding projected position operators.
Equivalently, one may use COM and relative coordinates,
\begin{equation}\nonumber
    R_\nu^W
    =
    \frac{x_{e,\nu}^W+x_{h,\nu}^W}{2},
    \qquad
    r_\nu^W
    =
    x_{h,\nu}^W-x_{e,\nu}^W,
    \label{eq:com-relative-wannier-data}
\end{equation}
where $R_\nu^W$ gives the COM location of the eWF, while $r_\nu^W$ gives its internal dipole.
We refer to $(R_\nu^W,r_\nu^W)$, or equivalently $(x_{e,\nu}^W,x_{h,\nu}^W)$, as the constituent-resolved exciton Wannier coordinates of $\ket{W_\nu}$.

\subsection{Position operators, non-Abelian Berry connections, and Quantum Geometric Dipoles}
\label{subsec:nonabelian-internal-geometry}

Let $\tBareXe$ and $\tBareXh$ denote the continuum electron and hole position operators projected onto the active exciton subspace. These projected constituent positions take the covariant form
\begin{equation}
    \tBareXe
    =
    \ii\partial_Q+\Ae(Q),
    \qquad
    \tBareXh
    =
    \ii\partial_Q+\Ah(Q),
    \label{eq:projected-constituent-positions}
\end{equation}
where $\Ae(Q)$ and $\Ah(Q)$ are the electron- and hole-resolved non-Abelian Berry connections. These connections generate electron- and hole-resolved non-Abelian Wilson lines and Wilson loops by path-ordered exponentiation, or equivalently by products of unitarized constituent Wilson-line elements on a finite momentum mesh (Appendix~\ref{app:exciton-wilson-loops}).

The average of the two constituent connections defines the equal-weight COM Berry connection,
\begin{equation}\nonumber
    \Abar(Q)
    =
    \frac{\Ae(Q)+\Ah(Q)}{2},
    \label{eq:com-berry-connection}
\end{equation}
so that the projected COM coordinate is
\begin{equation}\nonumber
    \tRop
    =
    \frac{\tBareXe+\tBareXh}{2}
    =
    \ii\partial_Q+\Abar(Q).
    \label{eq:projected-com-position}
\end{equation}
COM Berry geometry has been studied for individual exciton bands and for groups of exciton bands~\cite{yao2008,kwan2021,jankowski2025,haber2023,thompson2025,paiva2024}. 
The difference between the two constituent connections defines the QGD matrix,
\begin{equation}
    \Dmat(Q)
    =
    \Ah(Q)-\Ae(Q),
    \label{eq:relative-coordinate-matrix}
\end{equation}
so that the projected internal-dipole operator is
\begin{equation}
    \trop
    =
    \tBareXh-\tBareXe
    =
    \Dmat(Q).
    \label{eq:projected-relative-position}
\end{equation}
Thus $\Abar(Q)$ determines the COM geometry, while $\Dmat(Q)$ carries the internal-dipole geometry. For an isolated exciton band, $\Dmat(Q)$ reduces, up to the chosen physical-dipole sign convention, to the scalar gauge-invariant electron-hole connection difference discussed as an exciton polarization, dipole vector, shift vector, or QGD~\cite{cao2021,paiva2024,fertig2025,davenport2026berryology,davenport2026composite,mendez2026,chen2026,yang2026gianthelicaldipole,yang2026shift,hu2026shift}. Here, we retain the full constituent-resolved matrix $\Dmat(Q)$ within a chosen active exciton subspace. In a chosen active-band frame, its diagonal elements give the expectation values of the projected relative coordinate $\trop$ in the corresponding Bloch exciton states at fixed $Q$, while its off-diagonal elements encode interband QGD matrix elements.

The two objects behave differently under gauge transformations. Under a $Q$-dependent rotation of the active exciton frame, $\Abar(Q)$ transforms as a non-Abelian Berry connection, while $\Dmat(Q)$ transforms covariantly,
\begin{equation}
    \Dmat(Q)
    \rightarrow
    U^\dagger(Q)\Dmat(Q)U(Q).
    \label{eq:Dmat-covariant-transform}
\end{equation}
Consequently, the trace of $\Dmat(Q)$ gives the gauge-invariant total QGD at momentum $Q$, and its Brillouin-zone average gives the net internal dipole of the active exciton bands.
In the multiband case, however, the full matrix $\Dmat(Q)$ contains additional gauge-covariant internal-dipole information that is absent for a single isolated exciton band. This information can support localized eWFs carrying nonzero internal dipoles despite a vanishing net internal dipole.

The projected electron and hole positions need not commute. Their commutator is controlled by the covariant variation of the QGD matrix,
\begin{equation}
    [\tBareXe,\tBareXh]
    =
    \ii
    \left[
        \partial_Q\Dmat(Q)
        -
        \ii[\Abar(Q),\Dmat(Q)]
    \right].
    \label{eq:commutator-D-main}
\end{equation}
For a single band, the non-Abelian commutator term $[\Abar(Q),\Dmat(Q)]$ vanishes, so the projected electron and hole positions commute when the scalar QGD is $Q$ independent. In that case, the electron- and hole-localizing eWFs coincide~\cite{davenport2026berryology}. In the symmetry-broken isolated band example of Fig.~\ref{fig:figure1}, we numerically find that $\Dmat(Q)$ is dispersive and has a nonzero Brillouin-zone average (Fig.~\ref{fig:figure1}(e)). In the multiband case, by contrast, $Q$-independence of $\Dmat(Q)$ is not sufficient to make $[\tBareXe,\tBareXh]$ vanish, because the non-Abelian term $[\Abar(Q),\Dmat(Q)]$ may remain nonzero.

\subsection{Joint localization and the exciton spatial localizer}
\label{subsec:exciton-spatial-localizer}

Constituent-resolved eWFs require a localization procedure sensitive to both projected constituent positions $\tBareXe$ and $\tBareXh$. We quantify the quality of simultaneous constituent localization using the joint spread
\begin{equation} \nonumber
    \Omega_\nu
    =
    \Delta_\nu\tBareXe^2
    +
    \Delta_\nu\tBareXh^2,
    \label{eq:joint-constituent-spread}
\end{equation}
where
\begin{equation}\nonumber
    \Delta_\nu\hat{O}^2
    \equiv
    \bra{W_\nu}\hat{O}^2\ket{W_\nu}
    -
    \left(\bra{W_\nu}\hat{O}\ket{W_\nu}\right)^2,
    \label{eq:wannier-variance-definition}
\end{equation}
and we refer to eWFs that minimize $\sum_\nu\Omega_\nu$ as maximally localized.
Using $\tBareXe=\tRop-\trop/2$ and $\tBareXh=\tRop+\trop/2$, the same spread can be written as
\begin{equation}
    \Omega_\nu
    =
    2\Delta_\nu\tRop^2
    +
    \frac{1}{2}\Delta_\nu\trop^2.
    \label{eq:joint-com-relative-spread}
\end{equation}
We note that this work considers the joint spread \emph{within} the active exciton subspace. The overall physical spread of the eWF will additionally have a contribution arising from the quantum metric, which is gauge invariant with respect to the total variance of all eWFs and thus not susceptible to further minimization via Wannier constructions ~\cite{marzari1997,haber2023}.
For an isolated exciton band, both $r^W$ and the relative-coordinate variance $\Delta_\nu\trop^2$ are gauge invariant, so only the COM contribution $\Delta_\nu\tRop^2$ can be varied and minimized by the Wannier gauge.
Furthermore, the commutator~\eqref{eq:commutator-D-main} can be equivalently expressed in COM and relative coordinates as
\begin{equation}
    [\tBareXe,\tBareXh]
    =
    [\tRop,\trop],
    \label{eq:xe-xh-r-R-commutator}
\end{equation}
and thus the joint spread obeys~\cite{maccone_uncertainty}
\begin{equation}
    \Omega_\nu
    \geq
    \left|
    \bra{W_\nu}
    [\tBareXe,\tBareXh]
    \ket{W_\nu}
    \right|
    =
    \left|
    \bra{W_\nu}
    [\tRop,\trop]
    \ket{W_\nu}
    \right|.
    \label{eq:joint-spread-commutator-bound}
\end{equation}
A nonzero commutator rules out a common eigenbasis and places a lower bound on the joint spread of the exciton coordinates, equivalently in the constituent or COM/relative coordinates. This motivates the central contribution of this work: a real-space framework for constructing eWFs that treats projected constituent positions equally and maximally localizes them in both electron and hole coordinates.

We implement this framework by extending the spatial localizer approach, recently developed for non-interacting Wannier functions in crystalline insulators~\cite{gerhard2026}, to the interacting case. In the non-interacting 2D case, the spatial localizer approach yields ``intelligent states'' that are minimum uncertainty in position. Such states yield maximally localized Wannier functions in an effective model of $\mathrm{WSe}_2$~\cite{gerhard2026} without any ansatz or iterative optimization protocol. We claim the eWFs constructed using the spatial localizer approach to be maximally localized, in the sense of minimizing $\sum_\nu\Omega_\nu$. 

The continuum operators $\tBareXe$ and $\tBareXh$ encode the Berry connection geometry, while the finite projected periodic operators~\cite{davenport2026berryology} $\tXe$ and $\tXh$ provide the representatives used in numerical implementations. 
For trial electron and hole positions $x_e$ and $x_h$, the finite-size projected periodic position operators (Appendix~\ref{app:mc-position-shifts}) are decomposed into Hermitian cosine and sine components. We then define the exciton spatial localizer
\begin{equation}
\begin{aligned}
    \Lexc(x_e,x_h)
    &=
    \tXeC{x_e}\otimes\Gam{1}
    +
    \tXeS{x_e}\otimes\Gam{2}
    \\
    &\quad+
    \tXhC{x_h}\otimes\Gam{3}
    +
    \tXhS{x_h}\otimes\Gam{4},
\end{aligned}
\label{eq:exciton-localizer-main}
\end{equation}
where $\Gam{i}$ generate the Euclidean Clifford algebra $\mathrm{Cl}_{4,0}(\mathbb{R})$. The localizer combines the electron and hole position operator sine and cosine components into a single Hermitian operator compatible with periodic boundary conditions, and parametrized by the trial positions $x_e$ and $x_h$. 

The corresponding localizer indicator function (LIF) is
\begin{equation}
    \LIF(x_e,x_h)
    =
    \min
    |\sigma(\Lexc(x_e,x_h))|,
    \label{eq:exciton-lif-main}
\end{equation}
where $\sigma(\Lexc(x_e,x_h))$ denotes the spectrum of the localizer. We denote a LIF minimum by $(x_e^\star,x_h^\star)$. Such minima identify candidate Wannier coordinates of maximally localized eWFs along the electron and hole coordinates. Because these direct constituent coordinates and the COM/relative coordinates are related linearly, we equivalently parameterize the LIF as
\begin{equation}
\LIF(R,r)
\equiv
\LIF\left(x_e=R-\frac{r}{2},x_h=R+\frac{r}{2}\right).
\label{eq:lif-relative-coordinates}
\end{equation}
Figure~\ref{fig:figure1}(c) shows the LIF for the symmetry-broken isolated exciton band in Fig.~\ref{fig:figure1}(b). The minimum occurs at a nonzero relative coordinate $r^\star$, indicating the presence of a jointly localized state with nonvanishing internal electron-hole dipole. The minima in each unit cell (boundaries denoted by vertical gray lines) are periodic images of the same candidate COM coordinate rather than distinct eWFs.

At the candidate coordinate, we extract a localized state and perform $Q$-resolved L\"owdin orthogonalization to obtain eWFs. The Wannier coordinates $(x_{e,\nu}^W,x_{h,\nu}^W)$, or equivalently $(R_\nu^W,r_\nu^W)$, are calculated from the eWFs themselves. In general, the candidate $(R_\nu^\star,r_\nu^\star)$ need not coincide exactly with these evaluated centers. For every eWF constructed here, however, the LIF minima converge to the evaluated Wannier coordinates in the thermodynamic limit. At the finite sizes used, their differences are of order $10^{-4}$ in units of the lattice constant. Additionally, if $\comm{\tBareXe}{\tBareXh}=0$, the LIF minima and corresponding localized states extracted from the spatial localizer are precisely the Wannier coordinates and eWFs~\cite{gerhard2026}.

The electron-hole structure of each localized eWF can be visualized through the joint density
\begin{align}
    \rho(x_e,x_h) = \bra{W_\nu} n^{e}_{x_{e}}n^{h}_{x_{h}} \ket{W_\nu},
\end{align}
where $n^{e/h}_{x_{e/h}}$ measures the local conduction electron ($e$) or valence hole ($h$) density (Appendix~\ref{app:model}).
Figure~\ref{fig:figure1}(d) shows the joint density of the eWF extracted from the LIF minimum in Fig.~\ref{fig:figure1}(c). The nonzero internal dipole moment is evident from the density being centered away from the diagonal $x_e=x_h$. 

Figure~\ref{fig:figure1} therefore demonstrates the constituent-resolved construction in the minimal isolated band setting. The model parameterization deliberately breaks the symmetries that would force a scalar QGD to vanish. In Sec.~\ref{sec:multiband_excitons}, we return to the same bilayer model and show how restoring these symmetries instead constrains groups of eWFs through multiband symmetry orbits.

\section{Symmetry-Constrained Multiband Exciton Wannier Functions}
\label{sec:multiband_excitons}

We now apply the constituent-resolved framework of Sec.~\ref{sec:multiband-exciton-geometry} to exciton bands with various symmetries. The same bilayer SSH model used for the symmetry-broken single-band example in Fig.~\ref{fig:figure1} also supports symmetry-preserving two-band and six-band active exciton subspaces. The two-band example realizes a symmetry-constrained Wannier configuration with equal and opposite internal dipoles, while the six-band example demonstrates how the spatial localizer framework can also resolve the transverse interlayer dipole of eWFs. The complete Hamiltonian, numerical parameters, microscopic symmetry conditions, and symmetry derivations are given in Appendix~\ref{app:model}.

\subsection{Bilayer SSH model and symmetry regimes}

The model consists of two aligned SSH chains, with orbitals $A,B$ in the upper layer and $C,D$ in the lower layer. We choose the orbital embedding $x_A=x_C=-\frac14$, $x_B=x_D=\frac14$ and separate the two layers by the transverse distance $d_{\mathrm{layer}}$. Each layer contains alternating nearest-neighbor hoppings, while interlayer hybridization couples the two chains. A layer-dependent onsite potential $\pm\delta$ models the effect of a transverse electric field, and short-ranged intralayer and interlayer density-density interactions bind particle-hole excitations.

\begin{figure*}[!tb]
\centering
\includegraphics[width=1\linewidth]{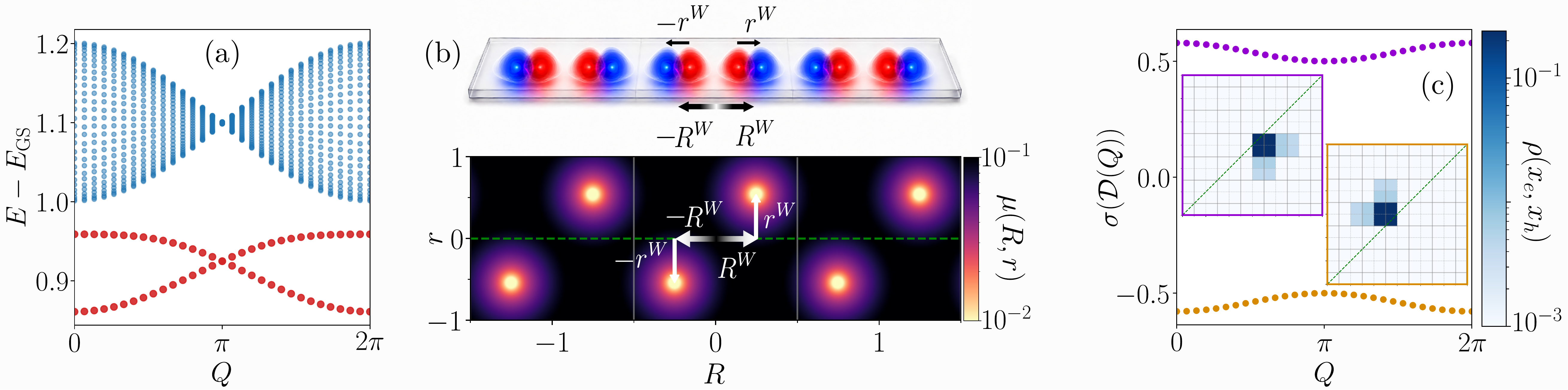}
\caption{{\bf Symmetry-constrained eWFs}
(a) Excitation spectrum in the single-exciton Hilbert space, with the active exciton bands highlighted in red.
(b) Graphical representation of the extracted Wannier configuration (top) and LIF (bottom). The LIF minima reveal a mirror-related two-point orbit of candidate coordinates $(R^\star,r^\star)$ and $(-R^\star,-r^\star)$ with $R^\star=1/4$, matching the corresponding evaluated Wannier coordinates.
The gray vertical lines denote unit-cell boundaries.
(c) Spectrum of the QGD matrix $\Dmat(Q)$, whose two eigenvalues form an equal-and-opposite pair, as independently required by $\mathcal{M}_x\mathcal{T}$ and $\mathcal{C}_{1/2}$. Insets: Joint densities of the two eWFs constructed from these candidate coordinates, whose evaluated centers obey $(R_2^W,r_2^W)=(-R_1^W,-r_1^W)$ up to lattice translations. The axes in the insets follow the axes in Fig.~\ref{fig:figure1}(d).
The green dashed lines in (b) and the insets of (c) denote $r=0$ ($x_e=x_h$).}
\label{fig:2}
\end{figure*}

The hopping and interaction terms can be chosen either to preserve or break mirror symmetry $\mathcal{M}_x$ and a nonsymmorphic Fock-space particle-hole symmetry $\mathcal{C}_{1/2}$. The latter interchanges creation and annihilation operators and is induced by a nonsymmorphic chiral spectral symmetry of the single-particle Hamiltonian that includes a translation by half a unit cell. We additionally consider the combined symmetry $\mathcal{S}_{1/2}=\mathcal{M}_x\mathcal{C}_{1/2}$.

The actions of the resulting exciton symmetries on $Q$ and the constituent-resolved coordinates are summarized in Table~\ref{tab:symmetry_transforms}. This table also indicates how each symmetry constrains the trace of the QGD matrix $\Dmat(Q)$. Among these constraints, $\mathcal{M}_x$ forces the net internal dipole to vanish by making $\Tr\Dmat(Q)$ odd in $Q$, while $\mathcal{C}_{1/2}$ and $\mathcal{M}_x\mathcal{T}$, where $\mathcal{T}$ is the spinless time reversal operator, force $\Tr\Dmat(Q)=0$ pointwise.

\begin{table}[t]
\centering
\begin{tabular}{|c|c|c|c|}
\hline
Symmetry & $Q\rightarrow$ & $(R,r)\rightarrow$ & $\Tr\Dmat(Q)$ \\
\hline
$\mathcal{T}$ & $-Q$ & $(R,r)$ & $\Tr\Dmat(Q) = \Tr\Dmat(-Q)$ \\
$\mathcal{M}_x$ & $-Q$ & $(-R,-r)$ & $\Tr\Dmat(Q) = -\Tr\Dmat(-Q)$  \\
$\mathcal{C}_{1/2}$ & $Q$ & $(R+1/2,-r)$ & $\Tr\Dmat(Q) = 0$  \\
$\mathcal{M}_x\mathcal{T}$ & $Q$ & $(-R,-r)$ & $\Tr\Dmat(Q) = 0$   \\
$\mathcal{S}_{1/2}$ & $-Q$ & $(1/2-R,r)$ & $\Tr\Dmat(Q) = \Tr\Dmat(-Q)$  \\
\hline
\end{tabular}
\caption{How symmetries transform the exciton momentum, $Q$, and the spatial coordinates, $(R,r)$, and how they constrain the trace of the QGD matrix $\Dmat(Q)$. All COM coordinates are defined modulo lattice translations. The considered symmetries are time reversal, $\mathcal{T}$, mirror, $\mathcal{M}_x$, nonsymmorphic particle-hole, $\mathcal{C}_{1/2}$, and the combined symmetries $\mathcal{M}_x\mathcal{T}$ and $\mathcal{S}_{1/2}=\mathcal{M}_x\mathcal{C}_{1/2}$.}
\label{tab:symmetry_transforms}
\end{table}

\begin{table}[b]
\centering
\begin{tabular}{|c|c|c|c|}
\hline
 & Regime 1 & Regime 2& Regime 3 \\
\hline
$N_\mathrm{exc}$ & 1 & 2 & 6 \\ 
$\delta$ & Strong & Strong & Weak \\
\hline
$\mathcal{T}$ & 1 & 1 & 1 \\
$\mathcal{M}_x$ & 0 & 1 & 1  \\
$\mathcal{C}_{1/2}$ & 0 & 1 & 1  \\
$\mathcal{M}_x\mathcal{T}$ & 0 & 1 & 1   \\
$\mathcal{S}_{1/2}$ & 0 & 1 & 1  \\
\hline
\end{tabular}
\caption{Number of active exciton bands, $N_\mathrm{exc}$, transverse field strength, $\delta$, and whether or not the corresponding system preserves each considered symmetry for the three regimes of Hamiltonian (Appendix~\ref{app:model}) considered in this work. Regimes 1, 2, and 3 respectively are presented in Figures~\ref{fig:figure1},~\ref{fig:2}, and~\ref{fig:3}. The symmetries are time reversal, $\mathcal{T}$, mirror, $\mathcal{M}_x$, nonsymmorphic particle-hole, $\mathcal{C}_{1/2}$, and the combined symmetries $\mathcal{M}_x\mathcal{T}$ and $\mathcal{S}_{1/2}=\mathcal{M}_x\mathcal{C}_{1/2}$. Number 1 (0) denotes the presence (absence) of symmetry.}
\label{tab:figure_info}
\end{table}

We consider three numerical regimes of the same microscopic model. Figure~\ref{fig:figure1} uses a strong $\delta$ parameterization with one isolated active exciton band. This regime preserves $\mathcal{T}$ but breaks $\mathcal{M}_x$, $\mathcal{C}_{1/2}$, $\mathcal{M}_x\mathcal{T}$, and $\mathcal{S}_{1/2}$, allowing the isolated band to support an eWF with nonzero internal dipole. Figure~\ref{fig:2} restores the broken symmetries in a strong $\delta$ regime with two active exciton bands. Figure~\ref{fig:3} uses a weaker $\delta$, for which both parent conduction bands and both parent valence bands contribute to a six-band active exciton subspace. This information is summarized in Table~\ref{tab:figure_info} and complete numerical parameters are collected in Table~\ref{tab:model_parameters} in Appendix~\ref{app:model}.

In the single-exciton Hilbert space, the mirror and nonsymmorphic particle-hole symmetries obey the relation
\begin{equation}
\mathcal{M}_x\mathcal{C}_{1/2}= T^{-1}_1 \mathcal{C}_{1/2}\mathcal{M}_x,
\end{equation}
where $T_1$ is a translation by one unit cell and $T_1(Q)=e^{-\ii Q}$. Therefore, we have the relation
\begin{equation}
\acomm{\mathcal{M}_x(Q=\pi)}{\mathcal{C}_{1/2}(Q=\pi)}=0.
\label{eq:symm_acomm}
\end{equation}
Thus, if an exciton Bloch state $\ket{\Psi_\pi}$ has mirror eigenvalue $\xi=\pm1$, then $\mathcal{C}_{1/2}\ket{\Psi_\pi}$ has the same energy and mirror eigenvalue $-\xi$. The active exciton bands must therefore be at least twofold degenerate at $Q=\pi$, although they need not be degenerate at generic $Q$.

\subsection{Symmetry-constrained two-band exciton Wannier functions}

Figure~\ref{fig:2}(a) shows the excitation spectrum in the symmetry-preserving, strong $\delta$ regime. Two bound exciton bands are separated from the particle-hole continuum and meet at the symmetry-enforced degeneracy at $Q=\pi$. The degenerate states have opposite mirror eigenvalues, as required by Eq.~\eqref{eq:symm_acomm}.

$\mathcal{M}_x$ acts as reflection about $R=0$, whereas $\mathcal{S}_{1/2}$ acts as reflection about $R=1/4$. These two reflections acting on the COM coordinate constrain the allowed Wannier configurations. Together they generate the generic Wannier coordinates orbit
\begin{equation}
(R^W,r^W), (-R^W,-r^W), (R^W+\frac12,-r^W), (\frac12-R^W,r^W).
\end{equation}
This four-point orbit reduces at special COM positions. In particular, $R=\pm1/4$ are individually fixed by the COM action of $\mathcal{S}_{1/2}$ and exchanged by $\mathcal{M}_x$. This allows the mirror-paired two-point orbit
\begin{align}
    (1/4,r^W),\qquad (-1/4,-r^W), \label{eq:mirror-related-orbit} 
\end{align}
while the corresponding symmetry-allowed Wannier coordinates that are fixed by $\mathcal{M}_x$ and exchanged by $\mathcal{C}_{1/2}$ and $\mathcal{S}_{1/2}$ are $(0,0),(1/2,0)$, with $r^W$ pinned to zero by $\mathcal{M}_x$.
The strong $\delta$ regime realizes the Wannier orbit~\eqref{eq:mirror-related-orbit}, seen in its LIF (Fig.~\ref{fig:2}(b)). The joint-density plots in the insets in Fig.~\ref{fig:2}(c) show the corresponding reversal of both the COM and relative coordinates.

The QGD spectrum in Fig.~\ref{fig:2}(c) demonstrates the nontrivial matrix-valued internal geometry of the active two-band exciton subspace. Both $\mathcal{C}_{1/2}$ and spinless $\mathcal{M}_x\mathcal{T}$ force the QGD eigenvalues to occur in equal and opposite pairs, i.e., $\Tr\Dmat(Q)=0$ pointwise. The nonzero eigenvalues nevertheless show that the full two-by-two matrix $\Dmat(Q)$ does not vanish. This nonzero, traceless internal structure allows the two eWFs to have equal and opposite internal dipoles. This possibility is intrinsically multiband, i.e., either $\mathcal{C}_{1/2}$ or spinless $\mathcal{M}_x\mathcal{T}$ would force $\Dmat(Q)=0$, and thus $[\tRop,\trop]=0$, for an isolated exciton band.

Figure~\ref{fig:2} shows a symmetry-constrained multiband Wannier configuration whose COM coordinates are associated with specific symmetry representations of the active exciton bands. The correspondence between inversion eigenvalues and symmetry-pinned Wannier centers is standard for inversion-symmetric bands~\cite{alexandradinata2014} and has been applied explicitly to isolated exciton bands~\cite{davenport2024,jankowski2025,davenport2026berryology}. Because $\mathcal{M}_x$ acts as $R\mapsto-R$ on the longitudinal one-dimensional coordinate, its symmetry representations obey the same correspondence, with the corresponding labels interpreted here as mirror eigenvalues. Here, the simultaneous presence of mirror symmetry $\mathcal{M}_x$ and the nonsymmorphic particle-hole symmetry $\mathcal{C}_{1/2}$, together with the restriction to an isolated two-band exciton subspace, excludes the generic four-point Wannier orbit and restricts the two eWFs to an allowed two-point orbit. The configuration $\{R_\nu^W\}=\{1/4,-1/4\}$ realized in Fig.~\ref{fig:2} has mirror eigenvalue irreps $\{+,-\}$ at both $Q=0$ and $Q=\pi$, whereas the alternative configuration $\{R_\nu^W\}=\{0,1/2\}$ has either $\{+,+\}$ or $\{-,-\}$ at $Q=0$ and $\{+,-\}$ at $Q=\pi$. The COM centers therefore distinguish the symmetry band representation, whereas the paired nonzero internal dipoles do not by themselves define a distinct symmetry-protected topological phase, since symmetry does not generically quantize $\abs{r^W}$.

\subsection{Resolving interlayer exciton polarization}

\begin{figure*}[!htb]
\centering
\includegraphics[width=1\linewidth]{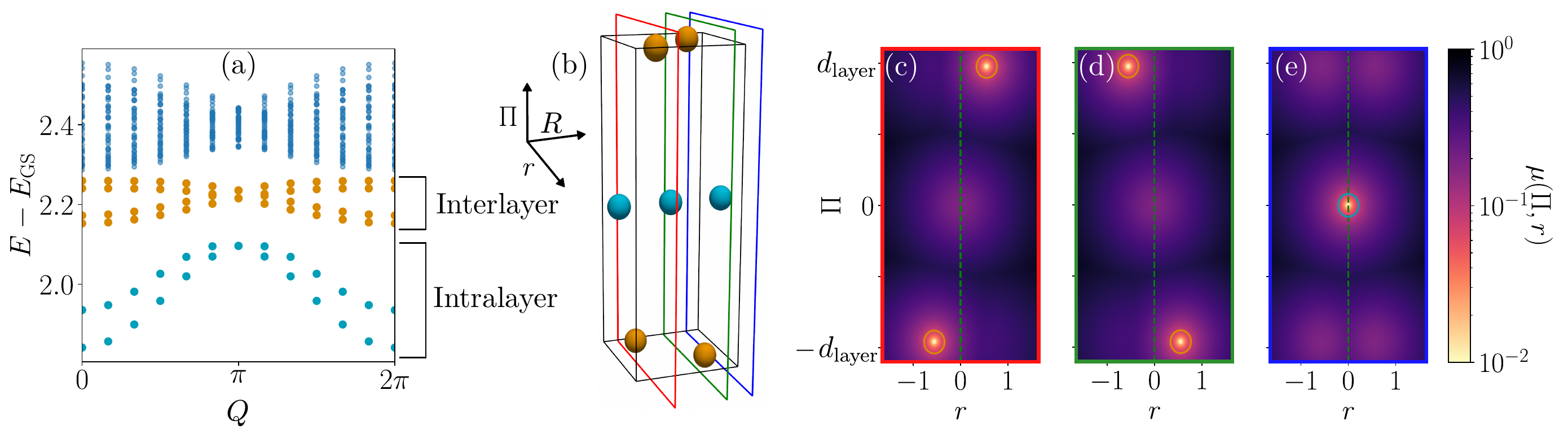}
\caption{{\bf Diagnosing interlayer exciton polarization.}
(a) Excitation spectrum, with the active exciton bands colored orange and light blue, respectively, for (predominantly) interlayer and intralayer exciton bands.
(b) Isosurface plot of $\mu(R,r,\Pi)=0.08$ for the six active exciton bands as a function of the COM coordinate, internal dipole, and interlayer dipole. The black wireframe corresponds to the bounds $R\in[-0.5,0.5],r\in[-1,1],$ and $\Pi\in[-d_{\mathrm{layer}},d_{\mathrm{layer}}]$. The red, green, and blue frames correspond to $r$-$\Pi$ LIF cuts at $R=-1/4$ (c), $R=1/4$ (d), and $R=1/2$ (e). The green dashed lines and orange/light blue contours in (c-e) respectively correspond to $r=0$ and the isosurface shown in (b).}
\label{fig:3}
\end{figure*}

The ``in-plane'' coordinates $(R,r)$ do not determine whether the electron and hole occupy the same or opposite layers. This distinction is essential in electrically tunable bilayers and multilayer exciton systems, where a transverse field can tune the hybridization between intralayer and interlayer exciton components and the resulting transverse dipoles strongly affect optical response, radiative lifetimes, transport, and interactions~\cite{barre2022opticalabsorption,jiang2021interlayer,tagarelli2023hybrid,schwandt2025ferroelectric}. We therefore extend the spatial localizer by including the exciton interlayer dipole as a third constituent-resolved coordinate.

To quantify the transverse positions of the two constituents, we define the electron and hole layer-position operators as
\begin{equation}
\Pi_{\mathrm{layer}}^{e/h}=\frac{d_{\mathrm{layer}}}{2}\sum_j\left(n_{j,A}^{e/h}+n_{j,B}^{e/h}-n_{j,C}^{e/h}-n_{j,D}^{e/h}\right),
\end{equation}
where the upper and lower layers are assigned transverse coordinates $+d_{\mathrm{layer}}/2$ and $-d_{\mathrm{layer}}/2$, respectively. In $\abs{e}=1$ units, their difference defines the exciton transverse dipole,
\begin{equation}
\Pi_{\mathrm{layer}}^{\mathrm{exc}}=\Pi_{\mathrm{layer}}^h-\Pi_{\mathrm{layer}}^e.
\end{equation}
Consequently, an intralayer exciton has $\Pi_{\mathrm{layer}}^{\mathrm{exc}}=0$, whereas the two orientations of an interlayer exciton have $\Pi_{\mathrm{layer}}^{\mathrm{exc}}=\pm d_{\mathrm{layer}}$. We note that, since interlayer polarization commutes with all considered symmetries and the two layers are coupled by interlayer hoppings, the interlayer dipole moment of eWFs is generally nonquantized.

With $\widetilde{\Pi}_{\mathrm{layer}}^{\mathrm{exc}}$ as the interlayer dipole operator projected onto the active exciton bands, we define the layer-resolved spatial localizer as
\begin{equation}
L_{\mathrm{exc}}^{\mathrm{layer}}(R,r,\Pi)=\Lexc(R,r)+\kappa_{\Pi}\left(\widetilde{\Pi}_{\mathrm{layer}}^{\mathrm{exc}}-\Pi I_{\mathrm{exc}}\right)\otimes\Gam{5},
\label{eq:interlayer_localizer}
\end{equation}
where $\Gam{5}$ is an additional Clifford generator, $I_{\mathrm{exc}}$ is an identity matrix on the active subspace, and $\kappa_{\Pi}\geq0$ is a relative weighting parameter (see Appendix~\ref{app:spatial-localizer}) that has units of inverse length. Furthermore, we have $\Lexc(R,r)\equiv\Lexc(x_e = R-r/2,x_h = R+r/2)$. The associated LIF is
\begin{equation}
\mu(R,r,\Pi)=\min\left|\sigma\left(L_{\mathrm{exc}}^{\mathrm{layer}}(R,r,\Pi)\right)\right|,
\end{equation}
so that a minimum of $\mu(R,r,\Pi)$ identifies a state jointly localized in its COM coordinate, in-plane relative coordinate, and transverse electron-hole dipole.

Figure~\ref{fig:3}(a) shows the excitation spectrum in the weak $\delta$ regime, where both parent conduction bands and both parent valence bands contribute. Six bound exciton bands are separated from the particle-hole continuum, with a lower group of two predominantly intralayer bands and an upper group of four predominantly interlayer bands. We use the union of all six bands as the active subspace for the layer-resolved localizer.

Since the LIF is now a scalar function of three variables, we use an isosurface for small $\mu(R,r,\Pi)$ to visualize the LIF (Fig.~\ref{fig:3}(b)) and reveal the constituent-resolved Wannier coordinates of the six-band subspace. To further elucidate the Wannier coordinates, we present three fixed-$R$ cuts through the $r$-$\Pi$ plane (Fig.~\ref{fig:3}(c-e)) corresponding to the colored frames in panel (b). The two unique minima near $\Pi=0$ correspond to predominantly intralayer eWFs, whereas the pairs of minima near $\Pi=\pm d_{\mathrm{layer}}$ correspond to the two orientations of predominantly interlayer eWFs. The LIF further reveals that the predominantly interlayer eWFs realize the configuration $\{(R_\nu^W,r_\nu^W)\}=\{(1/4,r^W),(-1/4,-r^W)\}$ at $\Pi\approx\pm d_{\mathrm{layer}}$, whereas the predominantly intralayer eWFs realize $\{(R_\nu^W,r_\nu^W)\}=\{(0,0),(1/2,0)\}$ at $\Pi\approx0$. We emphasize that, although there are three isosurfaces at $\Pi=0$, two of the isosurfaces are periodic images of one another.

This six-band example demonstrates that additional internal observables can be incorporated into the spatial localizer without altering the underlying constituent-resolved framework. Such extensions may be useful for diagnosing electrically controlled conversion between intralayer and interlayer excitons and for constructing localized descriptions of layer-polarized exciton transport and interactions.

\section{Discussion and Outlook}
\label{sec:discussion-conclusions}

We have developed a framework to build constituent-resolved eWFs in which each localized exciton retains the paired electron and hole centers, or equivalently the equal-weight COM and internal dipole coordinates. Projection onto an active exciton subspace gives constituent position operators $\tBareXe=\ii\partial_Q+\Ae(Q)$ and $\tBareXh=\ii\partial_Q+\Ah(Q)$ whose average determines the COM connection $\Abar(Q)$ and whose difference defines the non-Abelian QGD matrix $\Dmat(Q)=\Ah(Q)-\Ae(Q)$. The covariant $Q$ dependence of $\Dmat(Q)$ controls the noncommutativity of the projected constituent positions, which obstructs their simultaneous localization and bounds their joint spread from below. Whereas separate electron, hole, or COM Wilson loops diagnose the corresponding positions, they need not determine how the electron and hole centers are paired within an individual eWF. The exciton spatial localizer addresses this joint problem directly: it reframes the localization problem so that no gauge fixing is necessary, and yields eWFs jointly localized in both constituent coordinates by diagonalization alone. 

The bilayer SSH examples illustrate how this constituent-resolved structure changes between isolated and multiband exciton subspaces. In the symmetry-broken isolated band, the scalar QGD is $Q$ dependent, so the constituent positions do not commute, and has a nonzero Brillouin-zone average, so the eWF carries a nonzero internal dipole. When $\mathcal{M}_x$, $\mathcal{C}_{1/2}$, and spinless $\mathcal{M}_x\mathcal{T}$ are restored, the active subspace instead consists of two exciton bands that meet at a symmetry-enforced degeneracy at $Q=\pi$, resulting from the anticommutation of $\mathcal{M}_x$ and $\mathcal{C}_{1/2}$ at that momentum. Although $\mathcal{C}_{1/2}$ and $\mathcal{M}_x\mathcal{T}$ force the QGD spectrum to occur in equal and opposite pairs and hence require $\Tr\Dmat(Q)=0$ pointwise, they do not force the full QGD matrix to vanish. The spatial localizer resolves this nonzero traceless internal structure into the mirror-related Wannier configuration $(1/4,r^W)$ and $(-1/4,-r^W)$. Thus, a symmetry-preserving multiband exciton subspace can have zero net internal dipole while its individual eWFs carry equal and opposite internal dipoles. The COM configuration distinguishes the symmetry band representation of the active bands, whereas the paired nonzero internal dipoles do not by themselves define a distinct symmetry-protected topological phase, because symmetry does not generically quantize $\abs{r^W}$.

The six-band example further demonstrates that the spatial localizer can incorporate internal observables beyond the in-plane constituent coordinates. By adding the projected transverse layer dipole $\widetilde{\Pi}_{\mathrm{layer}}^{\mathrm{exc}}$ with an additional Clifford generator, we simultaneously resolve the COM coordinate, in-plane internal dipole, and intralayer or interlayer character of the eWFs. The resulting localizer distinguishes predominantly intralayer eWFs near $\Pi_{\mathrm{layer}}^{\mathrm{exc}}=0$ from the two orientations of predominantly interlayer eWFs near $\Pi_{\mathrm{layer}}^{\mathrm{exc}}=\pm d_{\mathrm{layer}}$. More generally, this construction provides a route for incorporating other physically relevant observables, such as spin or valley polarization, whenever they can be represented by projected operators within the active exciton subspace.

The present construction assumes an energetically isolated active exciton subspace, and at finite system size the candidate coordinates need not coincide exactly with the constituent centers evaluated from the resulting eWFs, although their differences vanish in the thermodynamic limit. Within these qualifications, constituent-resolved eWFs provide a natural basis for real-space models in which electric-field coupling, optical response, hopping, disorder, and exciton--exciton interactions depend on the internal electron-hole configuration. Natural extensions include higher-dimensional localizers for $(\mathbf{R},\mathbf{r})$, open-boundary constructions, and application of the present framework to first-principles Bethe--Salpeter exciton states, building on existing eWF and Wannier-decomposition approaches~\cite{haber2023,tao2025wfdx}. Similar constituent-resolved localizers may also be useful for other composite quasiparticles that admit localized bound-state descriptions.

\textit{Acknowledgments:} We thank Frank Schindler and Ajit Srivastava for discussions on this topic. This work was supported by the startup funds from Emory University and the Laboratory Directed Research and Development program at Sandia National Laboratories. Sandia National Laboratories is a multimission laboratory managed and operated by National Technology \& Engineering Solutions of Sandia, LLC, a wholly owned subsidiary of Honeywell International, Inc., for the U.S.\ DOE's National Nuclear Security Administration under contract DE-NA-0003525. The views expressed in the article do not necessarily represent the views of the U.S.\ DOE or the United States Government.
\newpage
\appendix

\onecolumngrid

\section{Single-Exciton Hilbert Space}
\label{app:multiband}
We begin by fixing the single-exciton Hilbert space and momentum convention used throughout the remaining appendices. The analytic derivations retain one parent conduction band and one parent valence band, while allowing an isolated group of $N_{\mathrm{exc}}$ active exciton bands.

We consider a one-dimensional periodic system of $\Lsize$ unit cells with the lattice constant set to unity, following the convention of the main text. The discrete momentum spacing is
\begin{equation}
    \dk = \frac{2\pi}{\Lsize}.
    \label{eq:mc-grid-step}
\end{equation}
All momentum labels are understood modulo a reciprocal lattice vector.
Let $c_k$ and $v_k$ annihilate electrons in one active conduction band and one active valence band, respectively, and let $\ket{u_{s,k}}$ denote the corresponding cell-periodic Bloch state for $s=c,v$. The reference state $\GS$ has the active valence band filled and the active conduction band empty.

A convenient basis for the single exciton (one conduction electron and one valence hole) sector at total exciton momentum $Q$ is
\begin{equation}
    \ket{Q,k}
    \equiv
    c_{Q+k}^{\dagger}v_k\GS.
    \label{eq:mc-ph-basis}
\end{equation}
The conduction electron momentum, physical hole momentum, and total momentum are related as
\begin{equation}
    q_e = Q+k,
    \qquad
    q_h = -k,
    \qquad
    Q = q_e + q_h.
    \label{eq:mc-momentum-bookkeeping}
\end{equation}
We retain $N_{\mathrm{exc}}$ active exciton bands and choose $N_{\mathrm{exc}}$ orthonormal exciton Bloch states that vary smoothly and periodically with total momentum,
\begin{equation}
    \ket{\Psi_{\alpha,Q}}
    =
    \sum_k \phi^Q_{\alpha,k}\,
    \ket{Q,k},
    \qquad \alpha=1,\ldots,N_{\mathrm{exc}},
    \label{eq:mc-exciton-state}
\end{equation}
with
\begin{equation}
    \sum_k \bar{\phi}^Q_{\alpha,k}\phi^Q_{\beta,k} = \delta_{\alpha\beta}.
    \label{eq:mc-exciton-orthonormality}
\end{equation}
Here $\alpha,\beta$ label the chosen exciton Bloch states, whereas $c,v$ label the single parent conduction and valence bands. If the active exciton bands are nondegenerate and do not cross, the states $\ket{\Psi_{\alpha,Q}}$ may be chosen as exciton energy eigenstates. More generally, when the bands cross or are treated together as a group, $\ket{\Psi_{\alpha,Q}}$ should be understood as a smooth mixture of energy eigenstates within the same active exciton subspace. We use sums over $k$ for a finite lattice and interpret them as normalized Brillouin-zone integrals in the thermodynamic limit. 

We define the $\nu$th eWF ($\nu \leq N_\mathrm{exc}$) of unit cell $j$ in the usual fashion~\cite{haber2023,marzari1997}, 
\begin{equation}
    \ket{W_{\nu,j}} = \frac{1}{\sqrt{L}}\sum_Q e^{-\ii Qj} \sum_{\alpha} \ket{\Psi_{\alpha,Q}} [U_W(Q)]_{\alpha \nu},
\end{equation}

and explicitly construct them via the projection method~\cite{marzari1997} and L\"owdin orthogonalization using trial states obtained from a spatial localizer (see Appendix~\ref{app:spatial-localizer}). We furthermore define $\ket{W_{\nu}}\equiv\ket{W_{\nu,j=0}}$ as the representative $\nu$th eWF. 

\section{Constituent Position Operators}
\label{app:mc-position-shifts}

In this appendix, we define the constituent periodic position operators that separately track the electron and hole positions in the single exciton, as in the construction of Ref.~\cite{davenport2026berryology}.
Given the conduction and valence band link factors
\begin{equation}
    S_c(k) = \braket{u_{c,k+\dk}}{u_{c,k}},
    \qquad
    S_v(k) = \braket{u_{v,k+\dk}}{u_{v,k}}.
    \label{eq:mc-electronic-links}
\end{equation}
The parent-band-projected many-body periodic position operators are
\begin{align}
    \Xe
    &=
    \sum_q S_c(q)\,
    c_{q+\dk}^{\dagger}c_q,
    \label{eq:mc-Ze-def}\\
    \Xh
    &=
    \sum_k S_v(k)\,
    v_k v_{k+\dk}^{\dagger}.
    \label{eq:mc-Zh-def}
\end{align}
\eqref{eq:mc-Ze-def} and \eqref{eq:mc-Zh-def} should not be confused with the many-body Resta exponential $\exp[\ii\dk\sum_j x_j\hat{n}_j]$~\cite{resta1998}, which contains terms that are higher order in the number of fermionic operators. 

Upon restriction to the sector containing exactly one conduction electron and one valence hole, $\Xe$ and $\Xh$ act as periodic position operators that shift the electron and hole degrees of freedom by a finite momentum step $\dk$. We note that the link factors $S_c(k)$ and $S_v(k)$ are understood to contain the intracell locations of the orbitals, as one can implement via either (i) the gauge choice of the Bloch Hamiltonian or (ii) an intracell position operator in the definition of the link factors.
Their actions on the basis \eqref{eq:mc-ph-basis} are
\begin{align}
    \Xe\ket{Q,k}
    &=
    S_c(Q+k)\ket{Q+\dk,k},
    \label{eq:mc-Ze-action}\\
    \Xh\ket{Q,k}
    &=
    S_v(k-\dk)\ket{Q+\dk,k-\dk},
    \label{eq:mc-Zh-action}
\end{align}
i.e.,
\begin{equation}
    \Xe:(Q,k)\mapsto(Q+\dk,k),
    \qquad
    \Xh:(Q,k)\mapsto(Q+\dk,k-\dk).
    \label{eq:mc-constituent-actions}
\end{equation}

This framework uses a common momentum mesh spacing for the single-particle momentum, $k$, and the many-body momentum, $Q$. By contrast, periodic implementations of a general weighted coordinate $R_\eta=(1-\eta)x_e + \eta x_h$ for $\eta\neq0,1$ can require more internal $k$ points than $Q$ points to achieve compatible constituent and exciton momentum meshes~\cite{haber2023}.

\section{Constituent Berry Connections and Wilson Loops}\label{app:bc_and_wilson_loops}
In this appendix, we derive the constituent Berry connections directly from expansions of finite link matrices for a subspace of active exciton bands. For completeness, we also present the resulting Wilson loops. For a single active exciton band, the Berry connection and Wilson loop construction reduces to that of Ref.~\cite{davenport2026berryology}. Here we retain the full matrix-valued links for an $N_{\mathrm{exc}}$-dimensional active exciton subspace.
\subsection{Electron-resolved link matrix and Berry connection}
\label{app:mc-electron}

Projecting $\Xe$ between neighboring total momenta defines the many-body links
\begin{equation}
    [\Me(Q)]_{\alpha\beta}
    \equiv
    \bra{\Psi_{\alpha,Q+\dk}}\Xe\ket{\Psi_{\beta,Q}}.
    \label{eq:mc-Me-def}
\end{equation}
Substituting \eqref{eq:mc-exciton-state} and \eqref{eq:mc-Ze-action} gives the finite-$\dk$ expression
\begin{equation}
    [\Me(Q)]_{\alpha\beta}
    =
    \sum_k
    \bar{\phi}^{Q+\dk}_{\alpha,k}\phi^Q_{\beta,k}
    S_c(Q+k).
    \label{eq:mc-Me-exact}
\end{equation}
Given the ordinary electronic conduction band Berry connection
\begin{equation}
    A_c(q)
    =
    \braket{u_{c,q}}{\ii\partial_q u_{c,q}},
    \label{eq:mc-Ac-elec}
\end{equation}
the small-$\dk$ expansions are
\begin{align}
    S_c(q)
    &= 1+\ii\dk A_c(q)+O(\dk^2),
    \label{eq:mc-Sc-expand}\\
    \bar{\phi}^{Q+\dk}_{\alpha,k}
    &= \bar{\phi}^Q_{\alpha,k}
      + \dk\partial_Q\bar{\phi}^Q_{\alpha,k}
      + O(\dk^2).
    \label{eq:mc-phi-p-expand}
\end{align}
Differentiating \eqref{eq:mc-exciton-orthonormality} yields
\begin{equation}
    \sum_k(\partial_Q\bar{\phi}^Q_{\alpha,k})\phi^Q_{\beta,k}
    =
    -\sum_k\bar{\phi}^Q_{\alpha,k}\partial_Q\phi^Q_{\beta,k}.
    \label{eq:mc-p-orthonormality}
\end{equation}
Consequently,
\begin{equation}
    \Me(Q) = I_{N_{\mathrm{exc}}} + \ii\dk\Ae(Q) + O(\dk^2),
    \label{eq:mc-Me-expand}
\end{equation}
where
\begin{equation}
    [\Ae(Q)]_{\alpha\beta}
    =
    \sum_k\left[
       \bar{\phi}^Q_{\alpha,k}\,\ii\partial_Q\phi^Q_{\beta,k}
       + \bar{\phi}^Q_{\alpha,k}\phi^Q_{\beta,k}A_c(Q+k)
    \right].
    \label{eq:mc-Ae-final}
\end{equation}
For $N_{\mathrm{exc}}=1$, \eqref{eq:mc-Ae-final} is the electron-localizing exciton Berry connection of Ref.~\cite{davenport2026berryology}.

\subsection{Hole-resolved link matrix and Berry connection}
\label{app:mc-hole}

Similarly, define
\begin{equation}
    [\Mh(Q)]_{\alpha\beta}
    \equiv
    \bra{\Psi_{\alpha,Q+\dk}}\Xh\ket{\Psi_{\beta,Q}}.
    \label{eq:mc-Mh-def}
\end{equation}
Using \eqref{eq:mc-Zh-action},
\begin{equation}
    [\Mh(Q)]_{\alpha\beta}
    =
    \sum_k
    \bar{\phi}^{Q+\dk}_{\alpha,k-\dk}
    \phi^Q_{\beta,k}
    S_v(k-\dk).
    \label{eq:mc-Mh-exact}
\end{equation}
The relevant expansions are
\begin{align}
    \bar{\phi}^{Q+\dk}_{\alpha,k-\dk}
    &=
    \bar{\phi}^Q_{\alpha,k}
    + \dk(\partial_Q-\partial_k)\bar{\phi}^Q_{\alpha,k}
    + O(\dk^2),
    \label{eq:mc-phi-hole-expand}\\
    S_v(k-\dk)
    &=
    \braket{u_{v,k}}{u_{v,k-\dk}}
    = 1 + \ii\dk A_v(k) + O(\dk^2),
    \label{eq:mc-Sv-expand}
\end{align}
where
\begin{equation}
    A_v(k) = \braket{u_{v,k}}{\ii\partial_k u_{v,k}}.
    \label{eq:mc-Av-elec}
\end{equation}
Using \eqref{eq:mc-exciton-orthonormality} and integration by parts,
\begin{equation}
    \sum_k[(\partial_Q-\partial_k)\bar{\phi}^Q_{\alpha,k}]\phi^Q_{\beta,k}
    =
    -\sum_k\bar{\phi}^Q_{\alpha,k}(\partial_Q-\partial_k)\phi^Q_{\beta,k}.
    \label{eq:mc-pk-integration-by-parts}
\end{equation}
It follows that
\begin{equation}
    \Mh(Q) = I_{N_{\mathrm{exc}}} + \ii\dk\Ah(Q) + O(\dk^2),
    \label{eq:mc-Mh-expand}
\end{equation}
with
\begin{equation}
    [\Ah(Q)]_{\alpha\beta}
    =
    \sum_k\left[
       \bar{\phi}^Q_{\alpha,k}\,\ii(\partial_Q-\partial_k)\phi^Q_{\beta,k}
       + \bar{\phi}^Q_{\alpha,k}\phi^Q_{\beta,k}A_v(k)
    \right].
    \label{eq:mc-Ah-final}
\end{equation}
The derivative $\partial_Q-\partial_k$ follows directly from the simultaneous shifts $Q\mapsto Q+\dk$ and $k\mapsto k-\dk$. For $N_{\mathrm{exc}}=1$, \eqref{eq:mc-Ah-final} is the hole-localizing exciton Berry connection
of Ref.~\cite{davenport2026berryology}.

\subsection{Unitarized links and Wilson loops}
\label{app:exciton-wilson-loops}

At finite momentum spacing, the overlap matrices $\Me(Q)$ and $\Mh(Q)$ need not be exactly unitary. Provided the relevant link is nonsingular, we define its unitary polar factor by
\begin{equation}
    F^s(Q)=M_s(Q)\left[M_s^\dagger(Q)M_s(Q)\right]^{-1/2},
    \qquad s=e,h.
    \label{eq:mc-unitarized-constituent-link}
\end{equation}
On a closed mesh, $Q_\ell=Q_0+\ell\dk$, the constituent Wilson loops are
\begin{equation}
    \mathcal{W}^s_{Q_0+2\pi\leftarrow Q_0}
    =\mathcal{P}\prod_{\ell=0}^{\Lsize-1}F^s(Q_\ell),
    \qquad s=e,h,
    \label{eq:mc-constituent-wilson-loops}
\end{equation}
or, in the thermodynamic limit,
\begin{equation}
    \mathcal{W}^e=\mathcal{P}\exp\left[\ii\oint dQ\,\Ae(Q)\right],
    \qquad
    \mathcal{W}^h=\mathcal{P}\exp\left[\ii\oint dQ\,\Ah(Q)\right],
    \label{eq:mc-continuum-constituent-wilson-loops}
\end{equation}
where $\mathcal{P}$ indicates the path ordering of the Wilson loop.
Their eigenphases and eigenstates yield electron- and hole-resolved Wannier centers and functions.
The average connection similarly generates the COM Wilson loop
\begin{equation}
    \mathcal{W}^{R}
    =\mathcal{P}\exp\left[\frac{\ii}{2}\oint dQ\,
    \bigl(\Ae(Q)+\Ah(Q)\bigr)\right],
    \label{eq:mc-com-wilson-loop}
\end{equation}
whose eigenphases and eigenstates yield COM-resolved Wannier centers and functions. We present a table comparing eWFs for the single band case constructed using the spatial localizer, $\mathcal{W}^e$, $\mathcal{W}^e$, and $\mathcal{W}^e$ in Table~\ref{tab:variance_compare}.

\section{Quantum Geometric Dipole and Projected-Position Algebra}\label{app:qgd_and_cont_pos_ops}
In this appendix, we derive a finite operator expression for the quantum geometric dipole (QGD) and expressions of the projected continuum position operators, $\tBareXe$ and $\tBareXh$, from the action of (projected) $X_e$ and $X_h$ on a smooth wavepacket in the active exciton subspace. We then discuss the commutator of the continuum projected position operators and the equivalence $[\tBareXe,\tBareXh] = [\tRop,\trop]$ presented in the main text.

\subsection{Relative-position block and the quantum geometric dipole}
\label{app:mc-internal}

The adjoint electron operator acts as
\begin{equation}
    \Xe^{\dagger}\ket{Q,k}
    =
    S_c^*(Q+k-\dk)\ket{Q-\dk,k}.
\end{equation}
Combining this with \eqref{eq:mc-Zh-action} gives
\begin{equation}
    \Xh\Xe^{\dagger}\ket{Q,k}
    =
    S_v(k-\dk)S_c^*(Q+k-\dk)\ket{Q,k-\dk}.
    \label{eq:mc-relative-action}
\end{equation}
The product is therefore block diagonal in total exciton momentum and shifts only the internal momentum coordinate.
The projected fixed-$Q$ block is
\begin{equation}
    [\Mrelmat(Q)]_{\alpha\beta}
    \equiv
    \bra{\Psi_{\alpha,Q}}\Xh\Xe^{\dagger}\ket{\Psi_{\beta,Q}}.
    \label{eq:mc-Rrel-def}
\end{equation}
Equation~\eqref{eq:mc-relative-action} yields
\begin{equation}
    [\Mrelmat(Q)]_{\alpha\beta}
    =
    \sum_k
    \bar{\phi}^Q_{\alpha,k-\dk}\phi^Q_{\beta,k}
    S_v(k-\dk)S_c^*(Q+k-\dk).
    \label{eq:mc-Rrel-exact}
\end{equation}
Using
\begin{align}
    \bar{\phi}^Q_{\alpha,k-\dk}
    &=
    \bar{\phi}^Q_{\alpha,k}
    - \dk\partial_k\bar{\phi}^Q_{\alpha,k}
    + O(\dk^2),\\
    S_v(k-\dk)S_c^*(Q+k-\dk)
    &=
    1 + \ii\dk\bigl[A_v(k)-A_c(Q+k)\bigr] + O(\dk^2),
\end{align}
and integrating by parts, one obtains
\begin{equation}
    \Mrelmat(Q) = I_{N_{\mathrm{exc}}} + \ii\dk\Dmat(Q) + O(\dk^2),
    \label{eq:mc-Rrel-expand}
\end{equation}
where
\begin{equation}
    [\Dmat(Q)]_{\alpha\beta}
    =
    \sum_k\left[
       -\bar{\phi}^Q_{\alpha,k}\,\ii\partial_k\phi^Q_{\beta,k}
       + \bar{\phi}^Q_{\alpha,k}\phi^Q_{\beta,k}
       \bigl(A_v(k)-A_c(Q+k)\bigr)
    \right]
    \label{eq:mc-D-final}
\end{equation}
is the (Hermitian) QGD. Thus $\Dmat(Q)$ can be diagonalized at each total momentum. Its eigenvalues are real relative-coordinate eigenvalues at fixed $Q$; the Wannier internal dipoles $r_\nu^W$ are obtained only after the full $Q$-space localization or Wannier construction.
To verify invariance under independent phase choices for the parent bands
\begin{equation}
    \ket{u_{c,q}}\longrightarrow e^{\ii\theta_c(q)}\ket{u_{c,q}},
    \qquad
    \ket{u_{v,k}}\longrightarrow e^{\ii\theta_v(k)}\ket{u_{v,k}}.
    \label{eq:mc-parent-gauge-transform}
\end{equation}
The exciton envelope coefficients must transform as
\begin{equation}
    \phi^Q_{\alpha,k}
    \longrightarrow
    \phi^{\prime Q}_{\alpha,k}
    =e^{-\ii\theta_c(Q+k)+\ii\theta_v(k)}\phi^Q_{\alpha,k},
    \label{eq:mc-envelope-parent-gauge}
\end{equation}
so that the physical exciton state in Eq.~\eqref{eq:mc-exciton-state} is unchanged, up to the irrelevant overall phase convention of the filled reference state. The parent-band connections transform according to
\begin{equation}
    A_c(q)\longrightarrow A_c(q)-\partial_q\theta_c(q),
    \qquad
    A_v(k)\longrightarrow A_v(k)-\partial_k\theta_v(k).
    \label{eq:mc-parent-connection-gauge}
\end{equation}
Meanwhile, the envelope-derivative term obeys
\begin{align}
    -\bar{\phi}^{\prime Q}_{\alpha,k}\,\ii\partial_k
    \phi^{\prime Q}_{\beta,k}
    ={}&
    -\bar{\phi}^Q_{\alpha,k}\,\ii\partial_k\phi^Q_{\beta,k}
    \nonumber\\
    &+\bar{\phi}^Q_{\alpha,k}\phi^Q_{\beta,k}
    \left[-\partial_k\theta_c(Q+k)+\partial_k\theta_v(k)\right].
    \label{eq:mc-envelope-derivative-gauge}
\end{align}
The second line of Eq.~\eqref{eq:mc-envelope-derivative-gauge} cancels exactly the phase-dependent change of $A_v(k)-A_c(Q+k)$ in Eq.~\eqref{eq:mc-D-final}. Thus every matrix element of $\Dmat(Q)$ is invariant under independent phase choices for the active conduction and valence parent bands.
Subtracting \eqref{eq:mc-Ae-final} from \eqref{eq:mc-Ah-final} gives
immediately
\begin{equation}
    \Dmat(Q) = \Ah(Q) - \Ae(Q).
    \label{eq:mc-D-difference}
\end{equation}
We refer to this Hermitian matrix as the \emph{quantum geometric dipole} (QGD). It is non-Abelian only because the active exciton subspace may have $N_{\mathrm{exc}}>1$.
For $N_{\mathrm{exc}}=1$, this reduces to the mean hole-electron dipole at fixed total momentum and therefore has the sign of the physical exciton polarization.
\eqref{eq:mc-D-final} is naturally interpreted in the thermodynamic limit as
\begin{equation}
    [\Dmat(Q)]_{\alpha\beta}
    =
    \bra{\Psi_{\alpha,Q}}\bigl(\tBareXh-\tBareXe\bigr)\ket{\Psi_{\beta,Q}},
    \label{eq:mc-D-position-matrix}
\end{equation}
with the understanding that $\tBareXh-\tBareXe$ can be obtained from the first-order expansion of the periodic operator $\Xh\Xe^{\dagger}$ (see Appendix~\ref{app:mc-commutator} for the derivation).
For $N_{\mathrm{exc}}=1$, Eq.~\eqref{eq:mc-D-final} reduces to the negative of the constituent-position difference derived in Ref.~\cite{davenport2026berryology}, owing solely to our physical-dipole sign convention. For $N_{\mathrm{exc}}\geq1$, the non-Abelian finite-link
comparison follows directly from the overlap construction above. 

\subsection{Projected positions and their commutator}
\label{app:mc-commutator}

We now connect the finite periodic position operators to the continuum projected position operators used in the main text. The key point is that $\Xe$ and $\Xh$ are finite periodic position operators that shift total momentum. Their first-order expansion in the momentum spacing $\dk$ gives the generators of projected constituent translations, which are the continuum projected position operators $\tBareXe$ and $\tBareXh$.

Let $\Pexc$ denote the projector onto the active exciton subspace, including all total momenta, and define the finite active-subspace operators
\begin{equation}
    \tXe=\Pexc\Xe\Pexc,
    \qquad
    \tXh=\Pexc\Xh\Pexc.
    \label{eq:mc-finite-active-position-operators}
\end{equation}
Using the link matrices defined in Eqs.~\eqref{eq:mc-Me-def} and \eqref{eq:mc-Mh-def}, the projected finite periodic position operators can be written as
\begin{align}
    \Pexc \Xe \Pexc
    &=
    \sum_Q
    \sum_{\alpha,\beta}
    \ket{\Psi_{\alpha,Q+\dk}}
    [\Me(Q)]_{\alpha\beta}
    \bra{\Psi_{\beta,Q}},
    \label{eq:mc-projected-Ze-link-form}\\
    \Pexc \Xh \Pexc
    &=
    \sum_Q
    \sum_{\alpha,\beta}
    \ket{\Psi_{\alpha,Q+\dk}}
    [\Mh(Q)]_{\alpha\beta}
    \bra{\Psi_{\beta,Q}}.
    \label{eq:mc-projected-Zh-link-form}
\end{align}
From Eqs.~\eqref{eq:mc-Me-expand} and \eqref{eq:mc-Mh-expand}, these link matrices have the small-$\dk$ expansions
\begin{align}
    \Me(Q)
    &=
    I_{N_{\mathrm{exc}}}
    +
    \ii\dk\Ae(Q)
    +
    O(\dk^2),
    \label{eq:mc-Me-expand-position-derivation}\\
    \Mh(Q)
    &=
    I_{N_{\mathrm{exc}}}
    +
    \ii\dk\Ah(Q)
    +
    O(\dk^2).
    \label{eq:mc-Mh-expand-position-derivation}
\end{align}
To extract the corresponding continuum operator, consider a smooth state in the active exciton subspace,
\begin{equation}
    \ket{\Phi}
    =
    \sum_Q
    \sum_{\beta}
    f_{\beta}(Q)\ket{\Psi_{\beta,Q}}.
    \label{eq:mc-active-wavepacket}
\end{equation}
Acting with $\Pexc\Xe\Pexc$ and reading off the coefficient of $\ket{\Psi_{\alpha,Q}}$ gives
\begin{equation}
    [\Pexc\Xe\Pexc f]_{\alpha}(Q)
    =
    \sum_{\beta}
    [\Me(Q-\dk)]_{\alpha\beta}
    f_{\beta}(Q-\dk).
    \label{eq:mc-Ze-action-on-envelope}
\end{equation}
Using Eq.~\eqref{eq:mc-Me-expand-position-derivation} and expanding the smooth envelope,
\begin{equation}
    f_{\beta}(Q-\dk)
    =
    f_{\beta}(Q)-\dk\,\partial_Q f_{\beta}(Q)+O(\dk^2),
    \label{eq:mc-envelope-shift-expand}
\end{equation}
we find
\begin{align}
    [\Pexc\Xe\Pexc f]_{\alpha}(Q)
    &=
    f_{\alpha}(Q)
    -
    \dk\,\partial_Q f_{\alpha}(Q)
    +
    \ii\dk
    \sum_{\beta}
    [\Ae(Q)]_{\alpha\beta}f_{\beta}(Q)
    +
    O(\dk^2)
    \nonumber\\
    &=
    \sum_{\beta}
    \left[
    \delta_{\alpha\beta}
    +
    \ii\dk
    \left(
    \ii\delta_{\alpha\beta}\partial_Q
    +
    [\Ae(Q)]_{\alpha\beta}
    \right)
    \right]
    f_{\beta}(Q)
    +
    O(\dk^2).
    \label{eq:mc-Ze-first-order-generator}
\end{align}
Thus, in the thermodynamic limit,
\begin{equation}
    \Pexc\Xe\Pexc
    =
    I_{\mathrm{exc}}+\ii\dk\,\tBareXe+O(\dk^2),
    \qquad
    \tBareXe
    =
    \ii\partial_Q+\Ae(Q).
    \label{eq:mc-Xe-generator}
\end{equation}
The same argument for the hole periodic position operator gives
\begin{equation}
    [\Pexc\Xh\Pexc f]_{\alpha}(Q)
    =
    \sum_{\beta}
    [\Mh(Q-\dk)]_{\alpha\beta}
    f_{\beta}(Q-\dk),
    \label{eq:mc-Zh-action-on-envelope}
\end{equation}
and therefore
\begin{equation}
    \Pexc\Xh\Pexc
    =
    I_{\mathrm{exc}}+\ii\dk\,\tBareXh+O(\dk^2),
    \qquad
    \tBareXh
    =
    \ii\partial_Q+\Ah(Q).
    \label{eq:mc-Xh-generator}
\end{equation}
Combining Eqs.~\eqref{eq:mc-Xe-generator} and \eqref{eq:mc-Xh-generator}, the continuum projected position operators act within the active exciton bands as
\begin{equation}
    \tBareXe = \ii\partial_Q + \Ae(Q),
    \qquad
    \tBareXh = \ii\partial_Q + \Ah(Q).
    \label{eq:mc-projected-positions}
\end{equation}
The common derivative term appears because both $\Xe$ and $\Xh$ advance the total exciton momentum by one grid step, as shown in Eq.~\eqref{eq:mc-constituent-actions}. The difference between electron and hole projected positions is therefore purely internal, i.e.,
\begin{equation}
    \tBareXh-\tBareXe
    =
    \Ah(Q)-\Ae(Q)
    =
    \Dmat(Q).
    \label{eq:mc-position-difference}
\end{equation}
Equivalently, this follows directly from the fixed-$Q$ relative block. Since Eq.~\eqref{eq:mc-Rrel-expand} gives
\begin{equation}
    \Mrelmat(Q)
    =
    I_{N_\mathrm{exc}}
    +
    \ii\dk\Dmat(Q)
    +
    O(\dk^2),
    \label{eq:mc-relative-block-generator}
\end{equation}
and $\Mrelmat(Q)$ is the active-subspace block of $\Xh\Xe^\dagger$, the first-order generator of the relative periodic position operator is precisely the projected relative coordinate $\tBareXh-\tBareXe=\Dmat(Q)$.

Defining
\begin{equation}
    \Abar(Q)=\frac{\Ae(Q)+\Ah(Q)}{2},
    \label{eq:mc-Abar-commutator-section}
\end{equation}
a direct calculation using Eq.~\eqref{eq:mc-projected-positions} gives
\begin{equation}
    [\tBareXe,\tBareXh]
    =
    \ii\left[
    \partial_Q\Dmat(Q)
    -
    \ii[\Abar(Q),\Dmat(Q)]
    \right].
    \label{eq:mc-position-commutator}
\end{equation}
Thus $[\tBareXe,\tBareXh]$ vanishes precisely when $\Dmat(Q)$ is covariantly constant with respect to the COM connection $\Abar(Q)$. In particular, commutation is not implied by a vanishing trace of $\Dmat(Q)$ or even by ordinary $Q$-independence of $\Dmat(Q)$ in a multiband active subspace since the non-Abelian term $[\Abar(Q),\Dmat(Q)]$ may remain nonzero.

The same noncommutativity can be written in COM and relative-coordinate variables. Given the projected COM and relative-coordinate operators
\begin{equation}
    \tRop
    =
    \frac{\tBareXe+\tBareXh}{2},
    \qquad
    \trop
    =
    \tBareXh-\tBareXe,
    \label{eq:mc-projected-R-r}
\end{equation}
or, equivalently,
\begin{equation}
    \tBareXe
    =
    \tRop-\frac{\trop}{2},
    \qquad
    \tBareXh
    =
    \tRop+\frac{\trop}{2}.
    \label{eq:mc-projected-Xe-Xh-from-R-r}
\end{equation}
A direct substitution yields
\begin{equation}
    [\tBareXe,\tBareXh]
    =
    [\tRop,\trop]
    =
    -[\trop,\tRop],
    \label{eq:mc-commutator-identity-relation}
\end{equation}
and thus
\begin{equation}
    [\tBareXe,\tBareXh]=0
    \quad\Longleftrightarrow\quad
    [\tRop,\trop]=0.
    \label{eq:mc-commutator-equivalence}
\end{equation}
Simultaneous localization of the projected electron and hole positions is therefore equivalent to simultaneous localization of the projected COM and relative coordinates. This is the operator-level reason why a constituent-resolved exciton Wannier description contains more information than a COM-only Wannier description.

\section{Symmetry Constraints on Wannier Coordinates and the Quantum Geometric Dipole}

In this section, we derive the constraints imposed by mirror symmetry $\mathcal{M}_x$ and spinless $\mathcal{M}_x\mathcal{T}$ on the QGD of a multiband active exciton subspace. We defer the constraint from the nonsymmorphic particle-hole symmetry $\mathcal{C}_{1/2}$ to Appendix~\ref{app:model}, where it can be presented in the context of the model used in this work.

Given a symmetry under which the active exciton bands are closed, its action on the active exciton Bloch states is encoded by an exciton sewing matrix, while its action on the fixed-$Q$ relative operator determines the constraint on $\Mrelmat(Q)$ and hence on $\Dmat(Q)$. This formulation keeps a single parent conduction and valence band, i.e., all sewing matrices in this section act in the $N_{\mathrm{exc}}$-dimensional active exciton subspace.

\subsection{$\mathcal{M}_x$ constraint}
\label{app:mc-mirror}

The action of $\mathcal{M}_x$ within the active exciton bands is described by the matrix $\BMx(Q)$ defined by
\begin{equation}
    \mathcal{M}_x\ket{\Psi_{\alpha,Q}}
    =
    \sum_\beta\ket{\Psi_{\beta,-Q}}[\BMx(Q)]_{\beta\alpha},
    \label{eq:mc-mirror-sewing}
\end{equation}
where $\BMx(Q)$ is unitary. The longitudinal relative position is mirror odd under $\mathcal{M}_x$, and at finite momentum spacing the corresponding periodic operator obeys
\begin{equation}
    \mathcal{M}_x\,\Xh\Xe^{\dagger}\,\mathcal{M}_x^{-1}
    =
    (\Xh\Xe^{\dagger})^{\dagger}.
    \label{eq:mc-mirror-relative-operator}
\end{equation}
Therefore the projected relative block satisfies
\begin{equation}
    \Mrelmat(Q)
    =
    \BMx^{\dagger}(Q)\,
    \Mrelmat^{\dagger}(-Q)\,
    \BMx(Q).
    \label{eq:mc-R-mirror}
\end{equation}
Expanding \eqref{eq:mc-R-mirror} to first order in $\dk$ gives
\begin{equation}
    \Dmat(Q)
    =
    -\BMx^{\dagger}(Q)\,
    \Dmat(-Q)\,
    \BMx(Q).
    \label{eq:mc-D-mirror}
\end{equation}
Taking the trace yields,
\begin{equation}
    \operatorname{Tr}\Dmat(Q) = -\operatorname{Tr}\Dmat(-Q),
    \label{eq:mc-trace-mirror}
\end{equation}
so the total Brillouin-zone-averaged internal dipole vanishes:
\begin{equation}
    \int_{\mathrm{BZ}}\frac{dQ}{2\pi}\,\operatorname{Tr}\Dmat(Q) = 0.
    \label{eq:mc-mirror-total-zero}
\end{equation}

However, the matrix $\Dmat(Q)$ need not vanish. For example, at a mirror-invariant momentum $Q_{\star}=0,\pi$, choosing active exciton states as eigenstates of $\mathcal{M}_x$ yields the sewing matrix,
\begin{equation}
    \BMx(Q_{\star})
    = \operatorname{diag}(\xi_1,\ldots,\xi_{N_{\mathrm{exc}}}),
    \qquad \xi_\alpha = \pm 1.
\end{equation}
Then
\begin{equation}
    [\Dmat(Q_{\star})]_{\alpha\beta}
    =
    -\xi_\alpha\xi_\beta[\Dmat(Q_{\star})]_{\alpha\beta}.
    \label{eq:mc-mirror-selection-rule}
\end{equation}
All matrix elements between states with equal mirror eigenvalues vanish, including every diagonal element, whereas matrix elements between states with opposite mirror eigenvalues are symmetry allowed.
For two active exciton bands with $\BMx = \sigma_z$,
\begin{equation}
    \Dmat(Q_{\star})
    = d_x(Q_{\star})\sigma_x + d_y(Q_{\star})\sigma_y,
    \qquad d_x,d_y\in\mathbb{R},
    \label{eq:mc-mirror-two-band}
\end{equation}
need not vanish despite having zero trace.

\subsection{Wannier-space internal-dipole pairing}
\label{app:mc-wannier-pairing}

Let $\ket{W_{\nu}}$ denote the eWFs for the home unit cell, constructed from a smooth Wannier gauge within the active exciton subspace. A mirror-symmetric set of eWFs may transform under $\mathcal{M}_x$ as
\begin{equation}
    \mathcal{M}_x\ket{W_{\nu}}
    =\ket{W_{\bar{\nu}}},
    \label{eq:mc-wannier-mirror-map}
\end{equation}
where $\nu=\bar{\nu}$ for eWFs mapped onto themselves by $\mathcal{M}_x$ (up to lattice translations), and $\nu\neq\bar{\nu}$ for two eWFs paired by $\mathcal{M}_x$. Since $\mathcal{M}_x$ reverses $\tBareXh-\tBareXe$, one finds
\begin{align}
    r_{\bar{\nu}}^W
    &=\bra{W_{\bar{\nu}}}
      (\tBareXh-\tBareXe)
      \ket{W_{\bar{\nu}}}\\
    &=-\bra{W_{\nu}}(\tBareXh-\tBareXe)\ket{W_{\nu}}
    =-r_\nu^W.
    \label{eq:mc-wannier-pairing}
\end{align}
An eWF mapped to itself therefore has zero internal dipole, whereas a mirror-exchanged pair carries equal and opposite internal dipoles.

\subsection{Spinless $\mathcal{M}_x\mathcal{T}$ constraint}
\label{app:mc-MxT}

Assume the active exciton bands and the parent conduction and valence bands are closed under spinless $\mathcal{M}_x\mathcal{T}$. $\mathcal{M}_x$ and time reversal each reverse the longitudinal momentum, so their product leaves $Q$ fixed. $\mathcal{M}_x$ reverses the longitudinal relative coordinate, while antiunitarity complex conjugates the periodic phase. The two sign reversals cancel, giving
\begin{equation}
    \mathcal{M}_x\mathcal{T}\,\Xh\Xe^{\dagger}\,
    (\mathcal{M}_x\mathcal{T})^{-1}=\Xh\Xe^{\dagger}.
    \label{eq:mc-MxT-relative-invariant}
\end{equation}
The antiunitary sewing matrix $B_{\mathcal{M}_x\mathcal{T}}(Q)$ is defined by
\begin{equation}
    \mathcal{M}_x\mathcal{T}\ket{\Psi_{\alpha,Q}}
    =\sum_\beta\ket{\Psi_{\beta,Q}}
    [B_{\mathcal{M}_x\mathcal{T}}(Q)]_{\beta\alpha}.
    \label{eq:mc-MxT-sewing}
\end{equation}
It follows that
\begin{equation}
    B_{\mathcal{M}_x\mathcal{T}}^{\dagger}(Q)\Mrelmat(Q)
    B_{\mathcal{M}_x\mathcal{T}}(Q)=\Mrelmat^*(Q),
    \label{eq:mc-R-MxT}
\end{equation}
and hence
\begin{equation}
    B_{\mathcal{M}_x\mathcal{T}}^{\dagger}(Q)\Dmat(Q)
    B_{\mathcal{M}_x\mathcal{T}}(Q)=-\Dmat^*(Q).
    \label{eq:mc-D-MxT}
\end{equation}
Since $\Dmat(Q)$ is Hermitian, its trace is real, and therefore
\begin{equation}
    \operatorname{Tr}\Dmat(Q)=0
    \label{eq:mc-MxT-trace-zero}
\end{equation}
pointwise in $Q$.

For spinless $(\mathcal{M}_x\mathcal{T})^2=+1$, the active states can locally be chosen so that $\mathcal{M}_x\mathcal{T}$ acts as complex conjugation. In this choice $B_{\mathcal{M}_x\mathcal{T}}(Q)=I$, and Eq.~\eqref{eq:mc-D-MxT} becomes
\begin{equation}
    \Dmat(Q)=-\Dmat^*(Q).
    \label{eq:mc-D-MxT-real-gauge}
\end{equation}
Together with Hermiticity, this implies
\begin{equation}
    \Dmat^T(Q) = -\Dmat(Q),
\end{equation}
leading to the result that every diagonal element vanishes, but off-diagonal matrix elements may survive for $N_{\mathrm{exc}}>1$. The spectrum is symmetric under $\lambda\mapsto-\lambda$ at each momentum. For $N_{\mathrm{exc}}=1$, Eq.~\eqref{eq:mc-D-MxT-real-gauge} forces $\Dmat(Q)=0$ pointwise, recovering the isolated band result of Ref.~\cite{davenport2026berryology}. For $N_{\mathrm{exc}}=2$, the most general allowed form is then
\begin{equation}
    \Dmat(Q) = d(Q)\sigma_y,
    \qquad d(Q)\in\mathbb{R},
    \label{eq:mc-MxT-two-band}
\end{equation}
with eigenvalues $\pm d(Q)$.

The analysis above retains one parent conduction band and one parent valence band for notational clarity. First-principles Bethe--Salpeter exciton states routinely carry multiple conduction- and valence-band indices~\cite{rohlfing2000,haber2023,tao2025wfdx}. If several parent bands are retained, the exciton envelope acquires the corresponding conduction- and valence-band indices, while $S_c$, $S_v$, and the scalar constituent sewing phases are replaced by matrices in the parent-band spaces. After constructing the constituent operators in this enlarged space, projection onto the chosen $N_{\mathrm{exc}}$-dimensional active exciton subspace and the spatial-localizer construction proceed as above.

\section{A Spatial Localizer for Excitons}
\label{app:spatial-localizer}
The commutator in Eq.~\eqref{eq:mc-position-commutator} quantifies the obstruction to simultaneous constituent localization. We now formulate the spatial localizer framework~\cite{gerhard2026} for excitons and, more generally, multiparticle bound states. The construction uses exciton periodic position operators, which act as finite-size density operators on a periodic system. All discussed bounds on the variance of states extracted from a spatial localizer follow directly from the general bound derived in Appendix C of Ref.~\cite{gerhard2026}. 

We note that, in the numerical implementation of a spatial localizer, projection onto the active subspace is done using the isometry or ``half projector'' $V_{\mathrm{exc}}$, which is a matrix with the active states as column vectors and we have $V_{\mathrm{exc}} V_{\mathrm{exc}}^\dagger = P_{\mathrm{exc}}$ and $V_{\mathrm{exc}}^\dagger V_{\mathrm{exc}}=I_{\mathrm{exc}}$. $V_{\mathrm{exc}}$ is analogous to $V_{\mathrm{occ}}$ discussed in Ref.~\cite{gerhard2026}. An operator $\hat{O}$ is then projected as $\tilde{O}=V_{\mathrm{exc}}^\dagger \hat{O}V_{\mathrm{exc}}$. The use of $V_{\mathrm{exc}}$ instead of $P_\mathrm{exc}$ removes the null space of $P_\mathrm{exc}$, which can result in spectral artifacts in a spatial localizer.

\subsection{Construction}

Given a set of $N_{\mathrm{exc}}$ energetically isolated, active exciton bands, we embed the finite active-subspace projected periodic position operators $\tXe,\tXh$ into generators of the (Euclidean) Clifford algebra $\mathrm{Cl}_{4,0}(\mathbb{R})$, yielding the \emph{excitonic spatial localizer}

\begin{equation}
    \Lexc(x_e,x_h)
    =\tXeC{x_e}\otimes\Gam{1}
    +\tXeS{x_e}\otimes\Gam{2}
    +\tXhC{x_h}\otimes\Gam{3}
    +\tXhS{x_h}\otimes\Gam{4},
    \label{eq:localizer}
\end{equation}
with
\begin{align}
    \tXeC{x_e}
    &=\operatorname{Re}[\tXe e^{-\ii\dk x_e}]-I_{\mathrm{exc}},\\
    \tXeS{x_e}
    &=\operatorname{Im}[\tXe e^{-\ii\dk x_e}],
\end{align}
where $I_{\mathrm{exc}}$ is the identity on the finite active exciton Hilbert space, $\Gam{\mu}$ are matrix representations (e.g., Weyl-Brauer matrices~\cite{brauer1935}) of $\mathrm{Cl}_{4,0}(\mathbb{R})$ generators, and the projected position operators $\tXe,\tXh$ are understood to have been made unitary, either by unitarizing the individual links or by a polar decomposition.

From $\Lexc(x_e,x_h)$, we consider the \emph{localizer indicator function} (LIF), 

\begin{align}
    \mu(x_e,x_h) = \min[\abs{\sigma(\Lexc(x_e,x_h))}],
\end{align}
where $\sigma(\hat{O})$ represents the spectrum of an operator $\hat{O}$.

\subsection{Extracting Localized States}\label{app:spatloc_extracting_states}
The LIF minima (zeros in the thermodynamic limit), denoted by $(x_e^\star,x_h^\star)$, of the LIF inform the centers of the localized representation. From these minima, we extract the localizer eigenstates $\ket{\psi^L(x_e^\star,x_h^\star)}\in\mathcal{H}_\text{active} \otimes \mathcal{H}_\Gamma$ where $\mathcal{H}_\text{active}, \mathcal{H}_\Gamma$ respectively are the Hilbert spaces on which the projected position operators and the Clifford generators act. To extract a localized state in $\mathcal{H}_{\text{active}}$, we perform a Schmidt decomposition of $\ket{\psi^L(x_e^\star,x_h^\star)}$,

\begin{equation}
    \ket{\psi^L(x_e^\star,x_h^\star)} = \sum_i s_i \left( \ket{\psch_i(x_e^\star,x_h^\star)} \otimes \ket{\esch_i(x_e^\star,x_h^\star)} \right),
    \label{eq:SchmidtDecomposition}
\end{equation} 
where $s_i \geq 0 $ are the Schmidt values, obeying $\sum_i s_i^2=1$, and $\ket{\psch_i(x_e^\star,x_h^\star)}$ and $\ket{\esch_i(x_e^\star,x_h^\star)}$ are respectively the physical and embedding Schmidt vectors. We order the Schmidt values $\{s_i\}$ in descending order, and the leading physical Schmidt vector, $\ket{\psch_1(x_e^\star,x_h^\star)}$, corresponds to the localized state.

In the thermodynamic limit, the spread of $\ket{\psch_1(x_e^\star,x_h^\star)}$ within the active subspace is bounded as
\begin{equation}\label{eq:app_text_variance_bound}
        \Delta_H^2 \leq \Delta\tBareXe^2 + \Delta\tBareXh^2 \leq \Delta_H^2+\satresidual,
\end{equation}
with
\begin{align}\label{eq:app_text_variance_bound_terms}
        &\Delta\tBareXe^2 = \expval{\tBareXe^2} - \expval{\tBareXe}^2 \nonumber \\
        &\Delta\tBareXh^2 = \expval{\tBareXh^2} - \expval{\tBareXh}^2 \nonumber \\
        &\Delta_H^2 = \abs{\expval{\comm{\tBareXe}{\tBareXh}}}, \nonumber   \\
        &\satresidual = \frac{\mu^2(x_e^\star,x_h^\star)}{\dk^2} - \left(\expval{\tBareXe - x_e^\star I_{\mathrm{exc}}}^2 + \expval{\tBareXh - x_h^\star I_{\mathrm{exc}}}^2\right) +  \sum_{i\neq 1} \frac{s_i}{s_1} \abs{\bra{\psch_1(x_e^\star,x_h^\star)} \comm{\tBareXe}{\tBareXh} \ket{\psch_i(x_e^\star,x_h^\star)}},
\end{align}
where all implicit expectation values are with respect to $\ket{\psch_1(x_e^\star,x_h^\star)}$.

\subsection{Bound Equivalence Between Direct and Relative Coordinates}

Equation~\eqref{eq:mc-commutator-identity-relation} shows that the two commutators $[\tBareXe,\tBareXh]$ and $[\tRop,\trop]$ are equivalent. Thus, it is natural to expect an equivalence between simultaneously localizing exciton wave functions in constituent and COM/relative coordinates. Here, we show that the bound~\eqref{eq:app_text_variance_bound} has an equivalent form in terms of the COM and relative coordinates $R,r$. 

We begin with the spread equivalence. By direct substitution, we have the following relations

\begin{align}
&\expval{\tBareXe^2}
= \expval{\left(\tRop - \frac{\trop}{2}\right)^2}  = \expval{\tRop^2} + \frac{1}{4}\expval{\trop^2} - \frac{1}{2} \expval{\acomm{\tRop}{\trop}}, \label{eq:app_rel_first}\\
&\expval{\tBareXh^2}
= \expval{\left(\tRop + \frac{\trop}{2}\right)^2}  = \expval{\tRop^2} + \frac{1}{4}\expval{\trop^2} + \frac{1}{2} \expval{\acomm{\tRop}{\trop}},\\
&\expval{\tBareXe}^2
= \expval{\left(\tRop - \frac{\trop}{2}\right)}^2  = \expval{\tRop}^2 + \frac{1}{4}\expval{\trop}^2 - \expval{\tRop}\expval{\trop},\\
&\expval{\tBareXh}^2
= \expval{\left(\tRop + \frac{\trop}{2}\right)}^2  = \expval{\tRop}^2 + \frac{1}{4}\expval{\trop}^2 + \expval{\tRop}\expval{\trop}, \\
&\expval{\tBareXe - x_e^\star I_{\mathrm{exc}}}^2 = \expval{\left(\tRop - \frac{\trop}{2}\right) - (R^\star - \frac{r^\star}{2}) I_{\mathrm{exc}}}^2 = \expval{\tRop - R^\star I_{\mathrm{exc}}}^2 + \frac{1}{4}\expval{\trop - r^\star I_{\mathrm{exc}}}^2 - \expval{\tRop - R^\star I_{\mathrm{exc}}}\expval{\trop - r^\star I_{\mathrm{exc}}},\\
&\expval{\tBareXh - x_h^\star I_{\mathrm{exc}}}^2 = \expval{\left(\tRop + \frac{\trop}{2}\right) - (R^\star + \frac{r^\star}{2}) I_{\mathrm{exc}}}^2 = \expval{\tRop - R^\star I_{\mathrm{exc}}}^2 + \frac{1}{4}\expval{\trop - r^\star I_{\mathrm{exc}}}^2 + \expval{\tRop - R^\star I_{\mathrm{exc}}}\expval{\trop - r^\star I_{\mathrm{exc}}},\label{eq:app_rel_last}
\end{align}
where $R^\star,r^\star$ are defined as

\begin{align}
    R^\star = \frac{x_e^\star + x_h^\star}{2}, \quad r^\star = x_h^\star - x_e^\star.
\end{align}
From the relations above [Equations (\ref{eq:app_rel_first}-\ref{eq:app_rel_last})], we then have the following equivalences,
\begin{align}
    &\Delta\tBareXe^2 + \Delta\tBareXh^2 = 2\left(\Delta\tRop^2 + \frac{1}{4}\Delta \trop^2\right), \\ 
    &\expval{\tBareXe - x_e^\star I_{\mathrm{exc}}}^2 + \expval{\tBareXh - x_h^\star I_{\mathrm{exc}}}^2 = 2\left(\expval{\tRop - R^\star I_{\mathrm{exc}}}^2 + \frac{1}{4}\expval{\trop - r^\star I_{\mathrm{exc}}}^2\right),
\end{align}
resulting in the alternate, but equivalent formulation of the terms in~\eqref{eq:app_text_variance_bound},
\begin{align}
    &\Delta_H^2 \leq 2\left(\Delta\tRop^2 + \frac{1}{4}\Delta \trop^2\right) \leq \Delta_H^2+\satresidual, \\ 
    &\Delta_H^2 = \abs{\expval{\comm{\tRop}{\trop}}}\\
    &\satresidual = \frac{\mu^2(R^\star,r^\star)}{\dk^2} - \left(2\expval{\tRop - R^\star I_{\mathrm{exc}}}^2 + \frac{1}{2}\expval{\trop - r^\star I_{\mathrm{exc}}}^2\right) +  \sum_{i\neq 1} \frac{s_i}{s_1} \abs{\bra{\psch_1(R^\star,r^\star)} \comm{\tRop}{\trop} \ket{\psch_i(R^\star,r^\star)}}.
\end{align}
The factors of $2$ and $\frac{1}{2}$ arise because the $R$ and $r$ coordinate directions are orthogonal but not normalized. An orthonormal version of such a coordinate transformation is
\begin{align}
    &(x_e,x_h)\rightarrow(R'=\sqrt{2}R,r'=\sqrt{2}r/2), \\
    &\mu(R',r') \equiv \mu(x_e = \frac{R' - r'}{\sqrt{2}},x_h = \frac{R' + r'}{\sqrt{2}})
\end{align}
which, at the cost of the natural physical interpretation of the $R,r$ coordinates, yields the relatively clean bound
\begin{align}
    &\Delta_H^2 \leq \Delta\tRop'^2 + \Delta \trop'^2 \leq \Delta_H^2+\satresidual, \\ 
    &\Delta_H^2 = \abs{\expval{\comm{\tRop'}{\trop'}}}\\
    &\satresidual = \frac{\mu^2(R'^\star,r'^\star)}{\dk^2} - \left(\expval{\tRop' - R'^\star I_{\mathrm{exc}}}^2 + \expval{\trop' - r'^\star I_{\mathrm{exc}}}^2\right) +  \sum_{i\neq 1} \frac{s_i}{s_1} \abs{\bra{\psch_1(R'^\star,r'^\star)} \comm{\tRop'}{\trop'} \ket{\psch_i(R'^\star,r'^\star)}}.
\end{align}

More generally, one can relate the variances of an initial set of coordinates $\hat{X}_j$ to a new set of coordinates $\hat{Y}_i=\sum_j M_{ij}\hat{X}_j$, where the rows $\boldsymbol m_i$ of $M$ form a complete orthogonal set with $\boldsymbol m_i\cdot\boldsymbol m_j=\ell_i^2\delta_{ij}$. The normalized coordinates $\hat{Y}_i/\ell_i$ are then related to the original coordinates by an orthogonal transformation. Invariance of the trace of the covariance matrix therefore gives
\begin{align}
\sum_j\Delta\hat{X}_j^2=\sum_i\frac{1}{\ell_i^2}\Delta\hat{Y}_i^2.
\end{align}
Thus, the 1D relations between direct and relative coordinate spreads considered here readily generalize to higher spatial dimensions.

\subsection{Including Interlayer Polarization}

For the extended localizer in Eq.~\eqref{eq:interlayer_localizer}, the general spatial-localizer bound takes the form (in the thermodynamic limit)
\begin{equation}\label{eq:app_text_variance_bound_interlayer}
        \frac{\Delta_H^2+ \frac{\kappa_{\Pi}}{\dk}\left[ \abs{\expval{\comm{\widetilde{\Pi}_{\mathrm{layer}}^{\mathrm{exc}}}{\tBareXe}}} + \abs{\expval{\comm{\widetilde{\Pi}_{\mathrm{layer}}^{\mathrm{exc}}}{\tBareXh}}}\right]}{2} \leq \Delta\tBareXe^2 + \Delta\tBareXh^2 + \frac{\kappa_{\Pi}^2}{\dk^2}\Delta(\widetilde{\Pi}_{\mathrm{layer}}^{\mathrm{exc}})^2  \leq \Delta_H^2 + \satresidual + \satresidual_{\Pi},
\end{equation}
where
\begin{align}\label{eq:app_text_variance_bound_terms_interlayer}
        \satresidual_{\Pi} &= -\frac{\kappa_{\Pi}^2}{\dk^2} \expval{\widetilde{\Pi}_{\mathrm{layer}}^{\mathrm{exc}} - \Pi^\star I_{\mathrm{exc}}}^2 + \frac{\kappa_{\Pi}}{\dk}\left[ \abs{\expval{\comm{\widetilde{\Pi}_{\mathrm{layer}}^{\mathrm{exc}}}{\tBareXe}}} + \abs{\expval{\comm{\widetilde{\Pi}_{\mathrm{layer}}^{\mathrm{exc}}}{\tBareXh}}}\right ]  \\ &+ \frac{\kappa_{\Pi}}{\dk}\left[\sum_{i\neq 1} \frac{s_i}{s_1} \abs{\bra{\psch_1(x_e^\star,x_h^\star,\Pi^\star)} \comm{\widetilde{\Pi}_{\mathrm{layer}}^{\mathrm{exc}}}{\tBareXe} \ket{\psch_i(x_e^\star,x_h^\star,\Pi^\star)}} + \sum_{i\neq 1} \frac{s_i}{s_1} \abs{\bra{\psch_1(x_e^\star,x_h^\star,\Pi^\star)} \comm{\widetilde{\Pi}_{\mathrm{layer}}^{\mathrm{exc}}}{\tBareXh} \ket{\psch_i(x_e^\star,x_h^\star,\Pi^\star)}} \right]. \nonumber
\end{align}
$\satresidual$, $\Delta_H^2$, and $\ket{\psch_i(x_e^\star,x_h^\star,\Pi^\star)}$ are defined as in Sec.~\ref{app:spatloc_extracting_states}, with the extended LIF and Schmidt vectors evaluated at $(x_e^\star,x_h^\star,\Pi^\star)$. Thus, a natural choice of weight is $\kappa_\Pi=\Delta_k$. For the six-band example exhibited here, we tested $\kappa_\Pi\in[\dk,1]$ and found the LIF minima to be unchanged. The actual numerics for Fig.~\ref{fig:3} use $\kappa_\Pi=1$ and $d_{\mathrm{layer}}=1$. For context, at the considered system size of $L=12$ unit cells, we have $\Delta_k\approx0.524$. We note that we use five of the six Clifford generators of $\mathrm{Cl}_{6,0}(\mathbb{R})$ in the numerical implementation.

\section{SSH Models and Microscopic Symmetries}\label{app:model}

\begin{figure}
    \centering
    \includegraphics[width=1\linewidth]{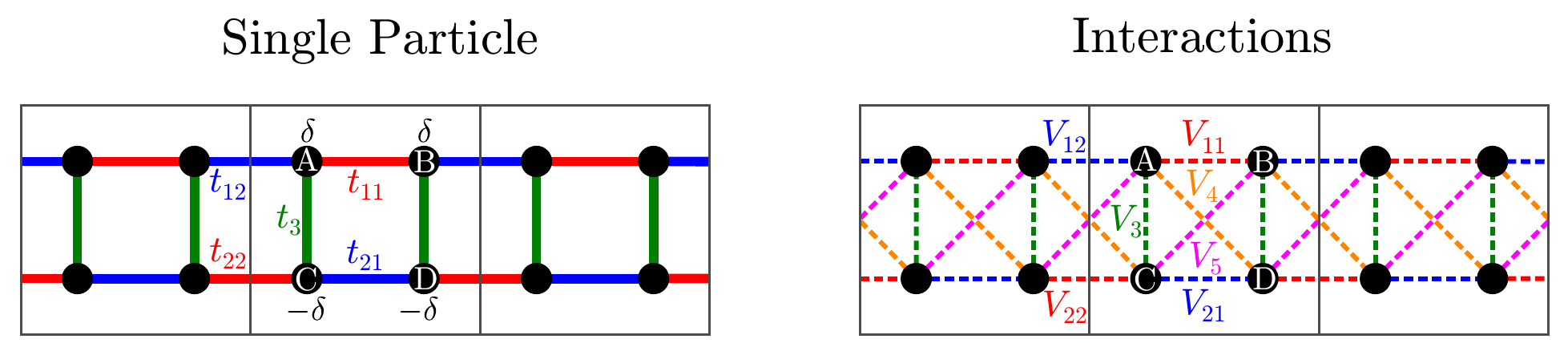}
    \caption{Graphical representation of the single-particle (left) and density-density interaction (right) terms in the bilayer SSH model used in this work.}
    \label{fig:model_diagrams}
\end{figure}

In this appendix, we present the models used in the numerical examples in the main text. Graphical representations of the single-particle and interacting Hamiltonians are presented in Fig.~\ref{fig:model_diagrams}, with numerical parameters presented in Table~\ref{tab:model_parameters}. All calculations are performed via exact diagonalization restricted to the single-exciton Hilbert space. In the large $\delta$ regime, we consider only one parent conduction and one parent valence band due to the energetic separation of the single-particle bands. We have verified that including the remote parent bands does not materially change the low-energy bound-state spectrum or the Wannier coordinates for the parameters used here.

\begin{table}[t]
\centering
\begin{tabular}{|c|c|c|c|c|c|c|c|c|c|c|c|c|c|c|c|}
\hline
 & $L$ & $N_{\mathrm{exc}}$ & $t_{11}$ & $t_{12}$ & $t_{21}$ & $t_{22}$ & $t_3$ & $\delta$ & $V_{11}$ & $V_{12}$ & $V_{21}$ & $V_{22}$ & $V_{3}$ & $V_{4}$ & $V_{5}$ \\ 
\hline
Symmetry broken & 30 & 1 & 1 & 0.1 & 0.05 & 1 & 0.025 & 1.5 & 0.4 & 0.2 & 0.2 & 0.4 & 0.3 & 0.2 & 0.1  \\ \hline
Symmetric, large $\delta$ & 30 & 2 & 1 & 0.05 & 0.05 & 1 & 0.025 & 1.5 & 0.4 & 0.2 & 0.2 & 0.4 & 0.3  & 0.2 & 0.2  \\ \hline
Symmetric, small $\delta$ & 12 & 6 & 1 & 0.05 & 0.05 & 1 & 0.025 & 0.005 & 0.4 & 0.2 & 0.2 & 0.4 & 0.3  & 0.2 & 0.2 \\ \hline
\end{tabular}
\caption{System sizes, number of active exciton bands, and model parameters used for the 3 different cases discussed in the main text. For the symmetry-broken phase, we implement $V_4\neq V_5$ and $t_{12}\neq t_{21}$ to break mirror symmetry $\mathcal{M}_x$, the particle--hole symmetry $\widehat{\mathcal{C}}_{1/2}$, and the combined symmetry $\widehat{\mathcal{S}}_{1/2}=\mathcal{M}_x\widehat{\mathcal{C}}_{1/2}$. For the $N_{\mathrm{exc}}=6$ example, where we implement a spatial localizer that considers interlayer polarization, we use $d_{\mathrm{layer}}=1$ and $\kappa_{\Pi}=1$.}
\label{tab:model_parameters}
\end{table}

\subsection{Single-particle Hamiltonian and orbital embedding}
We set the lattice constant to unity and label unit cells by $j\in\mathbb{Z}$. The bilayer SSH model has four orbitals per unit cell, ordered as $(A,B,C,D)$, with $A,B$ in the top layer and $C,D$ in the bottom layer. We choose the physical intracell coordinates
\begin{equation}
x_A^{\mathrm{orb}}=x_C^{\mathrm{orb}}=-\frac14,\qquad x_B^{\mathrm{orb}}=x_D^{\mathrm{orb}}=\frac14,
\label{eq:bilayer-orbital-embedding}
\end{equation}
so that
\begin{equation}
x_{j,\alpha}=j+x_\alpha^{\mathrm{orb}},\qquad \hat{x}\ket{j,\alpha}=x_{j,\alpha}\ket{j,\alpha}.
\label{eq:bilayer-physical-position}
\end{equation}
The two layers are separated by $d_{\mathrm{layer}}$ in the transverse direction.

A single SSH layer has the Hamiltonian (in the periodic gauge)
\begin{equation}
h_m(k)=\left(t_{m1}+t_{m2}\cos k\right)\sigma_x+t_{m2}\sin k\,\sigma_y=\begin{pmatrix}0&t_{m1}+t_{m2}e^{-ik}\\t_{m1}+t_{m2}e^{ik}&0\end{pmatrix},\qquad m=1,2.
\end{equation}
The single-particle Bloch Hamiltonian is
\begin{align}
    H_0(k)&=[(t_{11}+t_{12}\cos k)\sigma_x+t_{12}\sin k\,\sigma_y] \oplus [(t_{21}+t_{22}\cos k)\sigma_x+t_{22}\sin k\,\sigma_y] + \delta \tau_z \otimes \sigma_0 + t_3\tau_x \otimes \sigma_0 \nonumber \\
    &=\frac{\tau_0+\tau_z}{2}\otimes h_1(k)+\frac{\tau_0-\tau_z}{2}\otimes h_2(k)+\delta\,\tau_z\otimes\sigma_0+t_3\,\tau_x\otimes\sigma_0,
    \label{eq:bilayer-ssh-model-H0}
\end{align}
where $\sigma_i$ ($\tau_i$) denotes intralayer (interlayer) orbital DOF, and we present a graphical representation in Fig.~\ref{fig:model_diagrams}. 

The Hamiltonian~\eqref{eq:bilayer-ssh-model-H0} has the longitudinal mirror symmetry
\begin{equation}
    \mathcal{M}_x=\tau_0\otimes\sigma_x,\qquad \mathcal{M}_x H_0(k)\mathcal{M}_x^{-1}=H_0(-k),
\end{equation}
which reflects $x\mapsto-x$ within each chain while leaving the layer index unchanged.
When the hopping amplitudes satisfy $t_{11}=t_{22}$ and $t_{12}=t_{21}$, the Hamiltonian has an additional chiral spectral symmetry,
\begin{equation}
G_{1/2}(k)=\ii\,\tau_y\otimes\left[\begin{pmatrix}0&e^{-\ii k}\\1&0\end{pmatrix}\sigma_z\right]=\ii\,\tau_y\otimes\begin{pmatrix}0&-e^{-\ii k}\\1&0\end{pmatrix}.
\label{eq:half-shifted-chiral-operator}
\end{equation}
For arbitrary $\delta$ and $t_3$, we have
\begin{equation}
G_{1/2}(k)H_0(k)G_{1/2}^{\dagger}(k)=-H_0(k),\qquad G_{1/2}(k)^2=e^{-\ii k}I.
\end{equation}
Using the Fourier convention
\begin{equation}
\ket{k,\alpha}=\frac{1}{\sqrt{L}}\sum_j e^{\ii kj}\ket{j,\alpha},
\label{eq:periodic-bloch-convention}
\end{equation}
which is the periodic Bloch convention, the real-space action of $G_{1/2}$ is
\begin{align}
G_{1/2}\ket{j,A}&=-\ket{j,D},&
G_{1/2}\ket{j,B}&=\phantom{-}\ket{j+1,C},\nonumber\\
G_{1/2}\ket{j,C}&=\phantom{-}\ket{j,B},&
G_{1/2}\ket{j,D}&=-\ket{j+1,A}.
\label{eq:half-translation-real-space}
\end{align}
The corresponding orbital permutation $g$ is
\begin{align}
g(A_j)&=D_j,& g(B_j)&=C_{j+1},& g(C_j)&=B_j,& g(D_j)&=A_{j+1}.
\label{eq:glide-orbital-map}
\end{align}
Consequently,
\begin{equation}
G_{1/2}^2\ket{j,\alpha}=\ket{j+1,\alpha}.
\end{equation}
With the embedding in Eq.~\eqref{eq:bilayer-orbital-embedding}, the permutation $g$ translates every orbital position by one half lattice constant,
\begin{equation}
x_{g(j,\alpha)}=x_{j,\alpha}+\frac12.
\label{eq:glide-half-translation}
\end{equation}
Indeed, $A_j$ and $C_j$ at $j-1/4$ are mapped to $D_j$ and $B_j$ at $j+1/4$, while $B_j$ and $D_j$ at $j+1/4$ are mapped to $C_{j+1}$ and $A_{j+1}$ at $j+3/4$. Thus $G_{1/2}$ is a chiral spectral operator whose orbital action contains a geometric half translation and an exchange of the two layers. The real-space action of $\mathcal{M}_x$ is
\begin{align}
\mathcal{M}_x\ket{j,A}&=\ket{-j,B},& \mathcal{M}_x\ket{j,B}&=\ket{-j,A},\\
\mathcal{M}_x\ket{j,C}&=\ket{-j,D},& \mathcal{M}_x\ket{j,D}&=\ket{-j,C}.
\label{eq:bilayer-mirror-real-space}
\end{align}
The embedding makes this a geometric mirror reflection about $x=0$, since
\begin{equation}
x_{\mathcal{M}_x(j,\alpha)}=-x_{j,\alpha}.
\end{equation}
\subsection{Interacting Hamiltonian}

We consider short-ranged density-density interactions, namely nearest-neighbor (NN) intralayer and up to next-nearest-neighbor (NNN) interlayer interactions. The interacting Hamiltonian can be broken down into the top, bottom, and interlayer components,
\begin{align}\label{eq:bilayer-ssh-model-Hint}
    &H_{\mathrm{int}} = H^{AB} + H^{CD} + H^{\mathrm{inter}}, \\
    &H^{AB} = \sum_j V_{11}n_{j,A}n_{j,B} + V_{12}n_{j,B}n_{j+1,A}, \\
    &H^{CD} = \sum_j V_{21}n_{j,C}n_{j,D} + V_{22}n_{j,D}n_{j+1,C}, \\
    &H^{\mathrm{inter}} = \sum_j V_3 (n_{j,A}n_{j,C} +n_{j,B}n_{j,D}) 
    + V_{4} (n_{j,A}n_{j,D}+n_{j,B}n_{j+1,C})
    + V_{5} (n_{j,A}n_{j-1,D}+n_{j,B}n_{j,C}).
\end{align}

Within layers $m=1,2$, the couplings $V_{m1}$ and $V_{m2}$ denote the intracell and intercell density-density interactions, respectively. The coupling $V_3$ connects vertically aligned orbitals in opposite layers, while $V_4$ and $V_5$ connect right- and left-directed interlayer diagonals.

\subsection{Fock-space particle-hole symmetry induced by $G_{1/2}$}
\label{app:glide-particle-hole}

The anticommutation relation
\begin{equation}
G_{1/2}H_0G_{1/2}^{\dagger}=-H_0
\end{equation}
implies that $G_{1/2}$ maps positive-energy single-particle states to negative-energy states and conversely, i.e., a chiral symmetry~\cite{chiu2016}. However, $G_{1/2}$ is not itself a particle-hole transformation. Its ordinary second quantization maps electron creation operators to other electron creation operators and remains particle-number preserving. A particle-hole transformation is obtained by using $G_{1/2}$ as the orbital part of a canonical transformation that interchanges creation and annihilation operators~\cite{zirnbauer2021}.

Let $\ell=(j,\alpha)$ denote a combined unit-cell and orbital index, and let $h_0$ be the full real-space single-particle matrix. We define the Fock-space operator $\widehat{\mathcal{C}}_{1/2}$ by
\begin{equation}
\widehat{\mathcal{C}}_{1/2}c_\ell^\dagger\widehat{\mathcal{C}}_{1/2}^{-1}=\sum_m\left(G_{1/2}^{\dagger}\right)_{\ell m}c_m,\qquad \widehat{\mathcal{C}}_{1/2}c_\ell\widehat{\mathcal{C}}_{1/2}^{-1}=\sum_mc_m^\dagger\left(G_{1/2}\right)_{m\ell}.
\label{eq:fock-particle-hole-definition}
\end{equation}
Using the real-space action of $G_{1/2}$, this becomes
\begin{align}
\widehat{\mathcal{C}}_{1/2}c_{j,A}^{\dagger}\widehat{\mathcal{C}}_{1/2}^{-1}&=-c_{j,D},& \widehat{\mathcal{C}}_{1/2}c_{j,B}^{\dagger}\widehat{\mathcal{C}}_{1/2}^{-1}&=c_{j+1,C},\\
\widehat{\mathcal{C}}_{1/2}c_{j,C}^{\dagger}\widehat{\mathcal{C}}_{1/2}^{-1}&=c_{j,B},& \widehat{\mathcal{C}}_{1/2}c_{j,D}^{\dagger}\widehat{\mathcal{C}}_{1/2}^{-1}&=-c_{j+1,A}.
\label{eq:fock-particle-hole-real-space}
\end{align}
Thus $\widehat{\mathcal{C}}_{1/2}$ exchanges particles and holes while applying the non-onsite orbital permutation $g$, whose geometric action translates the physical position by one half lattice constant.
Acting on the quadratic Hamiltonian gives
\begin{equation}
\widehat{\mathcal{C}}_{1/2}H_0\widehat{\mathcal{C}}_{1/2}^{-1}=\operatorname{Tr}h_0-\sum_{\ell m}c_\ell^\dagger\left[G_{1/2}h_0^TG_{1/2}^{\dagger}\right]_{\ell m}c_m.
\label{eq:quadratic-ph-transformation}
\end{equation}
The exact particle-hole symmetry condition is therefore
\begin{equation}
G_{1/2}h_0^TG_{1/2}^{\dagger}=-h_0.
\label{eq:particle-hole-matrix-condition}
\end{equation}
In the localized basis used here, $h_0$ is real and symmetric. Equation~\eqref{eq:particle-hole-matrix-condition} therefore follows from the chiral anticommutation relation already established. Moreover, $\operatorname{Tr}h_0=0$, so
\begin{equation}
\widehat{\mathcal{C}}_{1/2}H_0\widehat{\mathcal{C}}_{1/2}^{-1}=H_0.
\end{equation}
The centered-density Hamiltonian introduced below preserves $\widehat{\mathcal{C}}_{1/2}$ when $t_{11}=t_{22}$ and $t_{12}=t_{21}$, together with the interaction conditions discussed in the next subsection. The uncentered interaction in Eq.~\eqref{eq:bilayer-ssh-model-Hint} has the same restriction to the half-filled sector, as shown below.

The particle-hole character is particularly transparent when $\abs{\delta}$ is large compared with the intralayer bandwidth and interlayer hybridization. For large positive $\delta$, the occupied states are predominantly localized in the $C,D$ layer while the empty states are predominantly localized in the $A,B$ layer. Equation~\eqref{eq:fock-particle-hole-real-space} then visibly exchanges the layer-polarized valence and conduction sectors. Large $\abs{\delta}$ is not required for the exact symmetry. Whenever the system is gapped at half filling, $G_{1/2}$ maps the complete negative-energy subspace to the complete positive-energy subspace. A large $\abs{\delta}$ only makes this conduction-valence correspondence approximately equivalent to a layer exchange and can help ensure that a restricted set of active conduction and valence bands is closed under the transformation.

\subsection{Interaction symmetry conditions}
\label{app:interaction-symmetries}

We now determine the conditions under which the density-density interaction in Eq.~\eqref{eq:bilayer-ssh-model-Hint} preserves mirror symmetry $\mathcal{M}_x$, the nonsymmorphic particle-hole symmetry, or their composition. The relevant orbital maps are given in Eqs.~\eqref{eq:bilayer-mirror-real-space} and \eqref{eq:glide-orbital-map}. The physical embedding does not alter the resulting bond permutations, and the signs in the real-space action of $G_{1/2}$ do not affect density operators.

It is useful to introduce the density measured relative to half filling,
\begin{equation}
\rho_{j,\alpha}=n_{j,\alpha}-\frac{1}{2},\qquad n_{j,\alpha}=c_{j,\alpha}^\dagger c_{j,\alpha}.
\end{equation}
For the combined index $\ell=(j,\alpha)$, $\mathcal{M}_x$ and the nonsymmorphic particle-hole transformation act on these densities as
\begin{equation}
\mathcal{M}_x\rho_\ell\mathcal{M}_x^{-1}=\rho_{\mathcal{M}_x(\ell)},\qquad \widehat{\mathcal{C}}_{1/2}\rho_\ell\widehat{\mathcal{C}}_{1/2}^{-1}=-\rho_{g(\ell)}.
\label{eq:centered-density-transformations}
\end{equation}
The two minus signs cancel in a density-density product, so the interaction symmetry conditions follow entirely from the permutation of the interacting bonds. The resulting coupling transformations are
\begin{center}
\begin{tabular}{|c|c|c|c|} \hline
coupling & $\mathcal{M}_x$ & $\widehat{\mathcal{C}}_{1/2}$ & $\mathcal{M}_x\widehat{\mathcal{C}}_{1/2}$\\
\hline
$V_{11}$ & $V_{11}$ & $V_{22}$ & $V_{22}$\\ \hline
$V_{12}$ & $V_{12}$ & $V_{21}$ & $V_{21}$\\ \hline 
$V_{21}$ & $V_{21}$ & $V_{12}$ & $V_{12}$\\ \hline 
$V_{22}$ & $V_{22}$ & $V_{11}$ & $V_{11}$\\ \hline 
$V_3$ & $V_3$ & $V_3$ & $V_3$\\ \hline 
$V_4$ & $V_5$ & $V_5$ & $V_4$\\ \hline 
$V_5$ & $V_4$ & $V_4$ & $V_5$ \\ \hline
\end{tabular}
\end{center}
For example, $\mathcal{M}_x$ maps the bond $n_{j,A}n_{j,D}$ into $n_{-j,B}n_{-j,C}$ and therefore exchanges the $V_4$ and $V_5$ interaction channels. The nonsymmorphic permutation maps $n_{j,A}n_{j,B}$ into $n_{j,D}n_{j+1,C}$ and $n_{j,B}n_{j+1,A}$ into $n_{j+1,C}n_{j+1,D}$, producing the exchanges $V_{11}\leftrightarrow V_{22}$ and $V_{12}\leftrightarrow V_{21}$.

The interaction is mirror symmetric under $\mathcal{M}_x$ when
\begin{equation}
\mathcal{M}_x:\qquad V_4=V_5.
\label{eq:interaction-mirror-condition}
\end{equation}
It is invariant under the nonsymmorphic particle-hole transformation when
\begin{equation}
\widehat{\mathcal{C}}_{1/2}:\qquad V_{11}=V_{22},\qquad V_{12}=V_{21},\qquad V_4=V_5.
\label{eq:interaction-particle-hole-condition}
\end{equation}
Finally, the interaction may preserve only the combined symmetry even when $\mathcal{M}_x$ and $\widehat{\mathcal{C}}_{1/2}$ are separately broken. The corresponding conditions are
\begin{equation}
\mathcal{M}_x\widehat{\mathcal{C}}_{1/2}:\qquad V_{11}=V_{22},\qquad V_{12}=V_{21},
\label{eq:interaction-combined-condition}
\end{equation}
with no requirement that $V_4=V_5$. When $V_4\neq V_5$, $\mathcal{M}_x$ and $\widehat{\mathcal{C}}_{1/2}$ are separately broken while their composition remains preserved. The coupling $V_3$ is unconstrained in all three cases because its two bond types are exchanged with one another.

Equation~\eqref{eq:centered-density-transformations} preserves particle-hole symmetry at the operator level when the interaction is written in terms of $\rho_\ell\rho_m$. The uncentered form in Eq.~\eqref{eq:bilayer-ssh-model-Hint} is equivalent within the half-filled single-exciton sector. Under either Eq.~\eqref{eq:interaction-particle-hole-condition} or Eq.~\eqref{eq:interaction-combined-condition}, the sum of the interaction strengths on all bonds incident on any orbital is the same,
\begin{equation}
\nu_V=V_{11}+V_{12}+V_3+V_4+V_5=V_{21}+V_{22}+V_3+V_4+V_5.
\label{eq:interaction-incident-sum}
\end{equation}
Consequently,
\begin{equation}
H_{\mathrm{int}}[n]=H_{\mathrm{int}}[\rho]+\frac{\nu_V}{2}\sum_\ell\rho_\ell+\mathrm{const.}
\label{eq:centered-uncentered-interaction}
\end{equation}
Since $\sum_\ell\rho_\ell=0$ at half filling, the centered and uncentered interactions differ only by an additive constant and generate the same dynamics in the single-exciton Hilbert space.

\subsection{Nonsymmorphic particle-hole symmetry and its composition with mirror symmetry}
\label{app:glide-exciton-action}

We next consider the unitary canonical particle-hole transformation $\widehat{\mathcal{C}}_{1/2}$ induced by the non-onsite orbital map $G_{1/2}$. We assume that the half-filled reference state is symmetry covariant, i.e.,
\begin{equation}
    \widehat{\mathcal{C}}_{1/2}\GS
    =e^{\ii\chi_{\mathcal{C}}}\GS.
\end{equation}
For a microscopic particle-hole bilinear $O_{\ell m}=c_\ell^\dagger c_m$, the real-space canonical transformation gives
\begin{align}
\widehat{\mathcal{C}}_{1/2}O_{\ell m}
\widehat{\mathcal{C}}_{1/2}^{-1}
&=\eta_\ell\eta_m^*\delta_{g(\ell),g(m)}
-\eta_\ell\eta_m^*c_{g(m)}^\dagger c_{g(\ell)},
\label{eq:exciton-ph-real-space}
\end{align}
where $\eta_m$ are factors resulting from the symmetry action, e.g., the signs in~\eqref{eq:half-translation-real-space}.
The contraction vanishes after projection to an interband excitation (e.g., restriction to the single-exciton Hilbert space). Thus, up to the displayed microscopic phases, the action of $\widehat{\mathcal{C}}_{1/2}$ on the single-exciton Hilbert space is
\begin{equation}
\widehat{\mathcal{C}}_{1/2} : \quad(e\text{ at }\ell,\;h\text{ at }m)
\mapsto
(e\text{ at }g(m),\;h\text{ at }g(\ell)).
\label{eq:exciton-endpoint-exchange}
\end{equation}

With one parent conduction band and one parent valence band, the action on the single-exciton basis contains only a scalar sewing phase,
\begin{equation}
\widehat{\mathcal{C}}_{1/2}\ket{Q,k}
=e^{\ii\chi_{\mathcal{C}}}u_{\mathcal{C}}(Q,k)
\ket{Q,-k-Q}.
\label{eq:exciton-glide-ph-momentum}
\end{equation}
The exchange of physical constituent momenta preserves $Q$, whereas $\mathcal{M}_x$ acts as
\begin{equation}
\mathcal{M}_x\ket{Q,k}=u_{\mathcal{M}_x}(Q,k)\ket{-Q,-k}.
\label{eq:exciton-mirror-momentum}
\end{equation}
Defining $\widehat{\mathcal{S}}_{1/2}=\mathcal{M}_x
\widehat{\mathcal{C}}_{1/2}$, one obtains
\begin{equation}
\widehat{\mathcal{S}}_{1/2}\ket{Q,k}
=e^{\ii\chi_{\mathcal{C}}}u_{\mathcal{S}}(Q,k)
\ket{-Q,k+Q}.
\label{eq:combined-exciton-momentum}
\end{equation}
From this point onward, we drop the hats and use $\mathcal{C}_{1/2}$ and $\mathcal{S}_{1/2}$ to denote the restrictions of the Fock-space canonical transformations $\widehat{\mathcal{C}}_{1/2}$ and $\widehat{\mathcal{S}}_{1/2}$ to the single-exciton Hilbert space. Using the geometric action $x_{g(\ell)}=x_\ell+1/2$ from Eq.~\eqref{eq:glide-half-translation}, Eq.~\eqref{eq:exciton-endpoint-exchange} implies
\begin{equation}
\mathcal{C}_{1/2}:(x_e,x_h)\mapsto
\left(x_h+\frac12,x_e+\frac12\right).
\label{eq:exciton-glide-constituent-action}
\end{equation}
Consequently, the COM and relative coordinates transform as
\begin{align}
{\mathcal{C}}_{1/2}:(R,r)&\mapsto
\left(R+\frac12,-r\right),
\label{eq:exciton-glide-coordinate-action}\\
{\mathcal{S}}_{1/2}:(R,r)&\mapsto
\left(\frac12-R,r\right).
\label{eq:combined-exciton-coordinate-action}
\end{align}
Accordingly, $\mathcal{C}_{1/2}$ pairs Wannier coordinates as $(R_\nu^W,r_\nu^W)\mapsto(R_\nu^W+1/2,-r_\nu^W)$, whereas $\mathcal{S}_{1/2}$ maps it to $(1/2-R_\nu^W,r_\nu^W)$. 

Projection onto the active exciton bands then promotes the scalar constituent phases in Eqs.~\eqref{eq:exciton-glide-ph-momentum} and \eqref{eq:combined-exciton-momentum} to sewing matrices $B_{\mathcal{C}}(Q)$ and $B_{\mathcal{S}}(Q)$, defined by
\begin{align}
\mathcal{C}_{1/2}\ket{\Psi_{\alpha,Q}}
&=\sum_\beta\ket{\Psi_{\beta,Q}}
[B_{\mathcal{C}}(Q)]_{\beta\alpha},\label{eq:mc-glide-sewing}\\
\mathcal{S}_{1/2}\ket{\Psi_{\alpha,Q}}
&=\sum_\beta\ket{\Psi_{\beta,-Q}}
[B_{\mathcal{S}}(Q)]_{\beta\alpha}.
\label{eq:mc-combined-sewing}
\end{align}
When the microscopic hopping and interaction conditions are satisfied, the corresponding exciton Hamiltonian obeys
\begin{equation}
\mathcal{C}_{1/2}(Q)H_{\mathrm{exc}}(Q)\mathcal{C}_{1/2}^\dagger(Q)
=H_{\mathrm{exc}}(Q),\qquad
\mathcal{S}_{1/2}(Q)H_{\mathrm{exc}}(Q)\mathcal{S}_{1/2}^\dagger(Q)
=H_{\mathrm{exc}}(-Q).
\label{eq:exciton-symmetry-covariance}
\end{equation}

Equation~\eqref{eq:exciton-glide-constituent-action} shows that the relative coordinate is odd under $\mathcal{C}_{1/2}$. Therefore,
\begin{equation}
\mathcal{C}_{1/2}\Xh\Xe^\dagger\mathcal{C}_{1/2}^{-1}
=(\Xh\Xe^\dagger)^\dagger.
\label{eq:mc-glide-relative-operator}
\end{equation}
Consequently,
\begin{equation}
\Mrelmat(Q)
 =B_{\mathcal{C}}^\dagger(Q)\Mrelmat^\dagger(Q)B_{\mathcal{C}}(Q).
\label{eq:mc-R-glide}
\end{equation}
Expanding to first order in $\dk$ yields
\begin{equation}
\Dmat(Q)
=-B_{\mathcal{C}}^\dagger(Q)\Dmat(Q)B_{\mathcal{C}}(Q).
\label{eq:mc-D-glide}
\end{equation}
Thus $\operatorname{Tr}\Dmat(Q)=0$ pointwise, the spectrum of $\Dmat(Q)$ is symmetric under $d\mapsto-d$, and an isolated exciton band has $\Dmat(Q)=0$ whenever $\mathcal{C}_{1/2}$ is preserved.

The combined symmetry leaves the relative coordinate invariant while mapping
$Q\mapsto-Q$. Thus, we have
\begin{equation}
\Mrelmat(Q)
 =B_{\mathcal{S}}^\dagger(Q)\Mrelmat(-Q)B_{\mathcal{S}}(Q).
\label{eq:mc-R-combined}
\end{equation}
Therefore
\begin{equation}
\Dmat(Q)
=B_{\mathcal{S}}^\dagger(Q)\Dmat(-Q)B_{\mathcal{S}}(Q).
\label{eq:mc-D-combined}
\end{equation}
Unlike $\mathcal{M}_x$ or $\mathcal{C}_{1/2}$ separately, this relation does not force a sign reversal of the internal dipole.

The nonsymmorphic algebra also manifests in the single-exciton Hilbert space.
The real-space transformation derived in
Eq.~\eqref{eq:fock-particle-hole-real-space} gives
\begin{equation}
\mathcal{C}_{1/2}(Q)^2=e^{-\ii Q},
\label{eq:exciton-glide-square}
\end{equation}
and $\mathcal{M}_x$ obeys
\begin{equation}
\mathcal{M}_x\mathcal{C}_{1/2}=T_1^{-1}\mathcal{C}_{1/2}\mathcal{M}_x.
\label{eq:exciton-mirror-glide-algebra}
\end{equation}
Thus, at $Q=\pi$,
\begin{equation}
\left\{\mathcal{M}_x(\pi),\mathcal{C}_{1/2}(\pi)\right\}=0.
\label{eq:exciton-zone-boundary-anticommutation}
\end{equation}
If both symmetries are preserved, this anticommutation enforces a twofold exciton degeneracy with opposite mirror eigenvalues. The combined symmetry alone satisfies $\mathcal{S}_{1/2}^2=1$ and does not enforce this degeneracy.

\subsection{Joint Density Calculations}
Here, we describe how the orbital-resolved joint densities in the main text figures are calculated.
Let $n^e_{j,\alpha}$ and $n^h_{j,\alpha}$ denote the local conduction-electron and valence-hole density operators projected onto the retained parent conduction and valence bands, respectively, where $j$ labels the unit cell and $\alpha\in\{A,B,C,D\}$ labels the orbital. Here, $n^h_{j,\alpha}$ measures the local absence of a valence electron. In the single-exciton sector, these operators are normalized such that
\begin{equation}
\sum_{j,\alpha}n^e_{j,\alpha}=\sum_{j,\alpha}n^h_{j,\alpha}=1,
\end{equation}
on the single-exciton sector.
Since the orbitals $A,C$ ($B,D$) have the same positions along the chain, we can then group them into the same local joint density operators as
\begin{align}
    n^{e/h}_{x_{e/h}} =\begin{cases}
    n^{e/h}_{j,A} + n^{e/h}_{j,C}, &  x_{e/h}=j-1/4, \\
    n^{e/h}_{j,B} + n^{e/h}_{j,D}, &  x_{e/h}=j+1/4.
\end{cases}
\end{align}
We then have the joint density distribution for an eWF $\ket{W_\nu}$ as
\begin{align}
    \rho(x_e,x_h) = \bra{W_\nu} n^{e}_{x_{e}}n^{h}_{x_{h}} \ket{W_\nu}.
\end{align}
From the joint density, we can calculate means and variances of an eWF. For example, mean position and total variance (including gauge-invariant contributions from the quantum metric~\cite{marzari1997,haber2023}) of a coordinate $q(x_e,x_h)$ can be calculated as 

\begin{align}
    \expval{q} &= \sum_{x_e,x_h} q(x_e,x_h) \rho(x_e,x_h),\\
    \Delta q^2 &= \sum_{x_e,x_h} (q(x_e,x_h) - \expval{q})^2 \rho(x_e,x_h).
\end{align}
Explicitly, we have the total variances 
\begin{align}
    \Delta {\BareXe}^2 &= \sum_{x_e,x_h} (x_e - \expval{x_e})^2\rho(x_e,x_h), \\
    \Delta {\BareXh}^2 &= \sum_{x_e,x_h} (x_h - \expval{x_h})^2\rho(x_e,x_h), \\
    \Delta {R}^2 &= \sum_{x_e,x_h} (R - \expval{R})^2\rho(x_e,x_h), \\
    \Delta {r}^2 &= \sum_{x_e,x_h} (r - \expval{r})^2\rho(x_e,x_h), \\
\end{align}
with the overall physical spread understood as $\Delta {\BareXe}^2 + \Delta {\BareXh}^2$.
We compare the variances of eWFs from the spatial localizer and Wilson loops in Table~\ref{tab:variance_compare} where one can see that the exciton spatial localizer yields eWFs with a smaller total variance than eWFs constructed from either electron- or hole-localizing Wilson loops, and identical variance of the COM-localizing Wilson loop eWF. The equivalence between the joint spreads of the WFs constructed using $\Lexc$ and $\mathcal{W}^R$ supports the claim that the spatial localizer approach provides maximally localized eWFs since, for a single isolated exciton band, the spread $\Delta r^2$ is gauge invariant.

\begin{table}[t]
\centering
\begin{tabular}{|c|c|c|c|c|} \hline
 & $L_{\mathrm{exc}}$ & $\mathcal{W}^R$ & $\mathcal{W}^e$ & $\mathcal{W}^h$\\
\hline
$\Delta {\BareXe}^2$ & 0.35053 & 0.35053 & 0.34972 & 0.35302\\ \hline
$\Delta {\BareXh}^2$ & 0.11607 & 0.11607 & 0.11861 & 0.11523\\ \hline 
$\Delta {R}^2$ & 0.12428 & 0.12428 & 0.12514 & 0.12511 \\ \hline 
$\Delta {r}^2$ & 0.43608 & 0.43608 & 0.43608 & 0.43608 \\ \hline 
$\Delta {\BareXe}^2 + \Delta {\BareXh}^2$ & 0.46659  & 0.46659 & 0.46833 & 0.46825 \\ \hline
\end{tabular}
\caption{Total spatial variances of eWFs, for the single band case, constructed using the exciton spatial localizer, $L_{\mathrm{exc}}$, and COM ($\mathcal{W}^R$), electron ($\mathcal{W}^e$) and hole ($\mathcal{W}^h$) localizing Wilson loops.}
\label{tab:variance_compare}
\end{table}

\bibliography{refs.bib}

\begin{thebibliography}{38}%
\makeatletter
\providecommand \@ifxundefined [1]{%
 \@ifx{#1\undefined}
}%
\providecommand \@ifnum [1]{%
 \ifnum #1\expandafter \@firstoftwo
 \else \expandafter \@secondoftwo
 \fi
}%
\providecommand \@ifx [1]{%
 \ifx #1\expandafter \@firstoftwo
 \else \expandafter \@secondoftwo
 \fi
}%
\providecommand \natexlab [1]{#1}%
\providecommand \enquote  [1]{``#1''}%
\providecommand \bibnamefont  [1]{#1}%
\providecommand \bibfnamefont [1]{#1}%
\providecommand \citenamefont [1]{#1}%
\providecommand \href@noop [0]{\@secondoftwo}%
\providecommand \href [0]{\begingroup \@sanitize@url \@href}%
\providecommand \@href[1]{\@@startlink{#1}\@@href}%
\providecommand \@@href[1]{\endgroup#1\@@endlink}%
\providecommand \@sanitize@url [0]{\catcode `\\12\catcode `\$12\catcode `\&12\catcode `\#12\catcode `\^12\catcode `\_12\catcode `\%12\relax}%
\providecommand \@@startlink[1]{}%
\providecommand \@@endlink[0]{}%
\providecommand \url  [0]{\begingroup\@sanitize@url \@url }%
\providecommand \@url [1]{\endgroup\@href {#1}{\urlprefix }}%
\providecommand \urlprefix  [0]{URL }%
\providecommand \Eprint [0]{\href }%
\providecommand \doibase [0]{https://doi.org/}%
\providecommand \selectlanguage [0]{\@gobble}%
\providecommand \bibinfo  [0]{\@secondoftwo}%
\providecommand \bibfield  [0]{\@secondoftwo}%
\providecommand \translation [1]{[#1]}%
\providecommand \BibitemOpen [0]{}%
\providecommand \bibitemStop [0]{}%
\providecommand \bibitemNoStop [0]{.\EOS\space}%
\providecommand \EOS [0]{\spacefactor3000\relax}%
\providecommand \BibitemShut  [1]{\csname bibitem#1\endcsname}%
\let\auto@bib@innerbib\@empty
\bibitem [{\citenamefont {Yao}\ and\ \citenamefont {Niu}(2008)}]{yao2008}%
  \BibitemOpen
  \bibfield  {author} {\bibinfo {author} {\bibfnamefont {W.}~\bibnamefont {Yao}}\ and\ \bibinfo {author} {\bibfnamefont {Q.}~\bibnamefont {Niu}},\ }\href {https://doi.org/10.1103/PhysRevLett.101.106401} {\bibfield  {journal} {\bibinfo  {journal} {Phys. Rev. Lett.}\ }\textbf {\bibinfo {volume} {101}},\ \bibinfo {pages} {106401} (\bibinfo {year} {2008})}\BibitemShut {NoStop}%
\bibitem [{\citenamefont {Srivastava}\ and\ \citenamefont {Imamo{\u{g}}lu}(2015)}]{srivastava2015}%
  \BibitemOpen
  \bibfield  {author} {\bibinfo {author} {\bibfnamefont {A.}~\bibnamefont {Srivastava}}\ and\ \bibinfo {author} {\bibfnamefont {A.}~\bibnamefont {Imamo{\u{g}}lu}},\ }\href {https://doi.org/10.1103/PhysRevLett.115.166802} {\bibfield  {journal} {\bibinfo  {journal} {Phys. Rev. Lett.}\ }\textbf {\bibinfo {volume} {115}},\ \bibinfo {pages} {166802} (\bibinfo {year} {2015})}\BibitemShut {NoStop}%
\bibitem [{\citenamefont {Zhou}\ \emph {et~al.}(2015)\citenamefont {Zhou}, \citenamefont {Shan}, \citenamefont {Yao},\ and\ \citenamefont {Xiao}}]{zhou2015}%
  \BibitemOpen
  \bibfield  {author} {\bibinfo {author} {\bibfnamefont {J.}~\bibnamefont {Zhou}}, \bibinfo {author} {\bibfnamefont {W.-Y.}\ \bibnamefont {Shan}}, \bibinfo {author} {\bibfnamefont {W.}~\bibnamefont {Yao}},\ and\ \bibinfo {author} {\bibfnamefont {D.}~\bibnamefont {Xiao}},\ }\href {https://doi.org/10.1103/PhysRevLett.115.166803} {\bibfield  {journal} {\bibinfo  {journal} {Phys. Rev. Lett.}\ }\textbf {\bibinfo {volume} {115}},\ \bibinfo {pages} {166803} (\bibinfo {year} {2015})}\BibitemShut {NoStop}%
\bibitem [{\citenamefont {Kwan}\ \emph {et~al.}(2021)\citenamefont {Kwan}, \citenamefont {Hu}, \citenamefont {Simon},\ and\ \citenamefont {Parameswaran}}]{kwan2021}%
  \BibitemOpen
  \bibfield  {author} {\bibinfo {author} {\bibfnamefont {Y.~H.}\ \bibnamefont {Kwan}}, \bibinfo {author} {\bibfnamefont {Y.}~\bibnamefont {Hu}}, \bibinfo {author} {\bibfnamefont {S.~H.}\ \bibnamefont {Simon}},\ and\ \bibinfo {author} {\bibfnamefont {S.~A.}\ \bibnamefont {Parameswaran}},\ }\href {https://doi.org/10.1103/PhysRevLett.126.137601} {\bibfield  {journal} {\bibinfo  {journal} {Phys. Rev. Lett.}\ }\textbf {\bibinfo {volume} {126}},\ \bibinfo {pages} {137601} (\bibinfo {year} {2021})}\BibitemShut {NoStop}%
\bibitem [{\citenamefont {Haber}\ \emph {et~al.}(2023)\citenamefont {Haber}, \citenamefont {Qiu}, \citenamefont {da~Jornada},\ and\ \citenamefont {Neaton}}]{haber2023}%
  \BibitemOpen
  \bibfield  {author} {\bibinfo {author} {\bibfnamefont {J.~B.}\ \bibnamefont {Haber}}, \bibinfo {author} {\bibfnamefont {D.~Y.}\ \bibnamefont {Qiu}}, \bibinfo {author} {\bibfnamefont {F.~H.}\ \bibnamefont {da~Jornada}},\ and\ \bibinfo {author} {\bibfnamefont {J.~B.}\ \bibnamefont {Neaton}},\ }\href {https://doi.org/10.1103/PhysRevB.108.125118} {\bibfield  {journal} {\bibinfo  {journal} {Phys. Rev. B}\ }\textbf {\bibinfo {volume} {108}},\ \bibinfo {pages} {125118} (\bibinfo {year} {2023})}\BibitemShut {NoStop}%
\bibitem [{\citenamefont {Davenport}\ \emph {et~al.}(2024)\citenamefont {Davenport}, \citenamefont {Knolle},\ and\ \citenamefont {Schindler}}]{davenport2024}%
  \BibitemOpen
  \bibfield  {author} {\bibinfo {author} {\bibfnamefont {H.}~\bibnamefont {Davenport}}, \bibinfo {author} {\bibfnamefont {J.}~\bibnamefont {Knolle}},\ and\ \bibinfo {author} {\bibfnamefont {F.}~\bibnamefont {Schindler}},\ }\href {https://doi.org/10.1103/PhysRevLett.133.176601} {\bibfield  {journal} {\bibinfo  {journal} {Phys. Rev. Lett.}\ }\textbf {\bibinfo {volume} {133}},\ \bibinfo {pages} {176601} (\bibinfo {year} {2024})}\BibitemShut {NoStop}%
\bibitem [{\citenamefont {Davenport}\ \emph {et~al.}(2026{\natexlab{a}})\citenamefont {Davenport}, \citenamefont {Knolle},\ and\ \citenamefont {Schindler}}]{davenport2026berryology}%
  \BibitemOpen
  \bibfield  {author} {\bibinfo {author} {\bibfnamefont {H.}~\bibnamefont {Davenport}}, \bibinfo {author} {\bibfnamefont {J.}~\bibnamefont {Knolle}},\ and\ \bibinfo {author} {\bibfnamefont {F.}~\bibnamefont {Schindler}},\ }\href {https://doi.org/10.1103/jtgq-vc7n} {\bibfield  {journal} {\bibinfo  {journal} {Phys. Rev. B}\ }\textbf {\bibinfo {volume} {113}},\ \bibinfo {pages} {045125} (\bibinfo {year} {2026}{\natexlab{a}})}\BibitemShut {NoStop}%
\bibitem [{\citenamefont {Wu}\ \emph {et~al.}(2017)\citenamefont {Wu}, \citenamefont {Lovorn},\ and\ \citenamefont {MacDonald}}]{wu2017}%
  \BibitemOpen
  \bibfield  {author} {\bibinfo {author} {\bibfnamefont {F.}~\bibnamefont {Wu}}, \bibinfo {author} {\bibfnamefont {T.}~\bibnamefont {Lovorn}},\ and\ \bibinfo {author} {\bibfnamefont {A.~H.}\ \bibnamefont {MacDonald}},\ }\href {https://doi.org/10.1103/PhysRevLett.118.147401} {\bibfield  {journal} {\bibinfo  {journal} {Phys. Rev. Lett.}\ }\textbf {\bibinfo {volume} {118}},\ \bibinfo {pages} {147401} (\bibinfo {year} {2017})}\BibitemShut {NoStop}%
\bibitem [{\citenamefont {Barr{\'e}}\ \emph {et~al.}(2022)\citenamefont {Barr{\'e}}, \citenamefont {Karni}, \citenamefont {Liu}, \citenamefont {O'Beirne}, \citenamefont {Chen}, \citenamefont {Ribeiro}, \citenamefont {Yu}, \citenamefont {Kim}, \citenamefont {Watanabe}, \citenamefont {Taniguchi}, \citenamefont {Barmak}, \citenamefont {Lui}, \citenamefont {Refaely-Abramson}, \citenamefont {da~Jornada},\ and\ \citenamefont {Heinz}}]{barre2022opticalabsorption}%
  \BibitemOpen
  \bibfield  {author} {\bibinfo {author} {\bibfnamefont {E.}~\bibnamefont {Barr{\'e}}}, \bibinfo {author} {\bibfnamefont {O.}~\bibnamefont {Karni}}, \bibinfo {author} {\bibfnamefont {E.}~\bibnamefont {Liu}}, \bibinfo {author} {\bibfnamefont {A.~L.}\ \bibnamefont {O'Beirne}}, \bibinfo {author} {\bibfnamefont {X.}~\bibnamefont {Chen}}, \bibinfo {author} {\bibfnamefont {H.~B.}\ \bibnamefont {Ribeiro}}, \bibinfo {author} {\bibfnamefont {L.}~\bibnamefont {Yu}}, \bibinfo {author} {\bibfnamefont {B.}~\bibnamefont {Kim}}, \bibinfo {author} {\bibfnamefont {K.}~\bibnamefont {Watanabe}}, \bibinfo {author} {\bibfnamefont {T.}~\bibnamefont {Taniguchi}}, \bibinfo {author} {\bibfnamefont {K.}~\bibnamefont {Barmak}}, \bibinfo {author} {\bibfnamefont {C.~H.}\ \bibnamefont {Lui}}, \bibinfo {author} {\bibfnamefont {S.}~\bibnamefont {Refaely-Abramson}}, \bibinfo {author} {\bibfnamefont {F.~H.}\ \bibnamefont {da~Jornada}},\ and\ \bibinfo {author} {\bibfnamefont {T.~F.}\ \bibnamefont {Heinz}},\ }\href
  {https://doi.org/10.1126/science.abm8511} {\bibfield  {journal} {\bibinfo  {journal} {Science}\ }\textbf {\bibinfo {volume} {376}},\ \bibinfo {pages} {406} (\bibinfo {year} {2022})}\BibitemShut {NoStop}%
\bibitem [{\citenamefont {Schwandt-Krause}\ \emph {et~al.}(2026)\citenamefont {Schwandt-Krause}, \citenamefont {Miloudi}, \citenamefont {Blundo}, \citenamefont {Deb}, \citenamefont {Heidkamp}, \citenamefont {Watanabe}, \citenamefont {Taniguchi}, \citenamefont {Schwartz}, \citenamefont {Stier}, \citenamefont {Finley}, \citenamefont {K{\"u}hn},\ and\ \citenamefont {Korn}}]{schwandt2025ferroelectric}%
  \BibitemOpen
  \bibfield  {author} {\bibinfo {author} {\bibfnamefont {J.}~\bibnamefont {Schwandt-Krause}}, \bibinfo {author} {\bibfnamefont {M.~E.~A.}\ \bibnamefont {Miloudi}}, \bibinfo {author} {\bibfnamefont {E.}~\bibnamefont {Blundo}}, \bibinfo {author} {\bibfnamefont {S.}~\bibnamefont {Deb}}, \bibinfo {author} {\bibfnamefont {J.-N.}\ \bibnamefont {Heidkamp}}, \bibinfo {author} {\bibfnamefont {K.}~\bibnamefont {Watanabe}}, \bibinfo {author} {\bibfnamefont {T.}~\bibnamefont {Taniguchi}}, \bibinfo {author} {\bibfnamefont {R.}~\bibnamefont {Schwartz}}, \bibinfo {author} {\bibfnamefont {A.}~\bibnamefont {Stier}}, \bibinfo {author} {\bibfnamefont {J.~J.}\ \bibnamefont {Finley}}, \bibinfo {author} {\bibfnamefont {O.}~\bibnamefont {K{\"u}hn}},\ and\ \bibinfo {author} {\bibfnamefont {T.}~\bibnamefont {Korn}},\ }\href {https://doi.org/10.1021/acs.nanolett.5c04932} {\bibfield  {journal} {\bibinfo  {journal} {Nano Letters}\ }\textbf {\bibinfo {volume} {26}},\ \bibinfo {pages} {214} (\bibinfo {year} {2026})}\BibitemShut {NoStop}%
\bibitem [{\citenamefont {Tagarelli}\ \emph {et~al.}(2023)\citenamefont {Tagarelli}, \citenamefont {Lopriore}, \citenamefont {Erkensten}, \citenamefont {Perea-Caus{\'i}n}, \citenamefont {Brem}, \citenamefont {Hagel}, \citenamefont {Sun}, \citenamefont {Pasquale}, \citenamefont {Watanabe}, \citenamefont {Taniguchi}, \citenamefont {Malic},\ and\ \citenamefont {Kis}}]{tagarelli2023hybrid}%
  \BibitemOpen
  \bibfield  {author} {\bibinfo {author} {\bibfnamefont {F.}~\bibnamefont {Tagarelli}}, \bibinfo {author} {\bibfnamefont {E.}~\bibnamefont {Lopriore}}, \bibinfo {author} {\bibfnamefont {D.}~\bibnamefont {Erkensten}}, \bibinfo {author} {\bibfnamefont {R.}~\bibnamefont {Perea-Caus{\'i}n}}, \bibinfo {author} {\bibfnamefont {S.}~\bibnamefont {Brem}}, \bibinfo {author} {\bibfnamefont {J.}~\bibnamefont {Hagel}}, \bibinfo {author} {\bibfnamefont {Z.}~\bibnamefont {Sun}}, \bibinfo {author} {\bibfnamefont {G.}~\bibnamefont {Pasquale}}, \bibinfo {author} {\bibfnamefont {K.}~\bibnamefont {Watanabe}}, \bibinfo {author} {\bibfnamefont {T.}~\bibnamefont {Taniguchi}}, \bibinfo {author} {\bibfnamefont {E.}~\bibnamefont {Malic}},\ and\ \bibinfo {author} {\bibfnamefont {A.}~\bibnamefont {Kis}},\ }\href {https://doi.org/10.1038/s41566-023-01198-w} {\bibfield  {journal} {\bibinfo  {journal} {Nat. Photon.}\ }\textbf {\bibinfo {volume} {17}},\ \bibinfo {pages} {615} (\bibinfo {year} {2023})}\BibitemShut {NoStop}%
\bibitem [{\citenamefont {Jiang}\ \emph {et~al.}(2021)\citenamefont {Jiang}, \citenamefont {Chen}, \citenamefont {Zheng}, \citenamefont {Zheng},\ and\ \citenamefont {Pan}}]{jiang2021interlayer}%
  \BibitemOpen
  \bibfield  {author} {\bibinfo {author} {\bibfnamefont {Y.}~\bibnamefont {Jiang}}, \bibinfo {author} {\bibfnamefont {S.}~\bibnamefont {Chen}}, \bibinfo {author} {\bibfnamefont {W.}~\bibnamefont {Zheng}}, \bibinfo {author} {\bibfnamefont {B.}~\bibnamefont {Zheng}},\ and\ \bibinfo {author} {\bibfnamefont {A.}~\bibnamefont {Pan}},\ }\href {https://doi.org/10.1038/s41377-021-00500-1} {\bibfield  {journal} {\bibinfo  {journal} {Light: Science \& Applications}\ }\textbf {\bibinfo {volume} {10}},\ \bibinfo {pages} {72} (\bibinfo {year} {2021})}\BibitemShut {NoStop}%
\bibitem [{\citenamefont {Wu}\ \emph {et~al.}(2018)\citenamefont {Wu}, \citenamefont {Lovorn},\ and\ \citenamefont {MacDonald}}]{wu2018opticalabsorption}%
  \BibitemOpen
  \bibfield  {author} {\bibinfo {author} {\bibfnamefont {F.}~\bibnamefont {Wu}}, \bibinfo {author} {\bibfnamefont {T.}~\bibnamefont {Lovorn}},\ and\ \bibinfo {author} {\bibfnamefont {A.~H.}\ \bibnamefont {MacDonald}},\ }\href {https://doi.org/10.1103/PhysRevB.97.035306} {\bibfield  {journal} {\bibinfo  {journal} {Physical Review B}\ }\textbf {\bibinfo {volume} {97}},\ \bibinfo {pages} {035306} (\bibinfo {year} {2018})}\BibitemShut {NoStop}%
\bibitem [{\citenamefont {G{\"o}tting}\ \emph {et~al.}(2022)\citenamefont {G{\"o}tting}, \citenamefont {Lohof},\ and\ \citenamefont {Gies}}]{gotting2022moire}%
  \BibitemOpen
  \bibfield  {author} {\bibinfo {author} {\bibfnamefont {N.}~\bibnamefont {G{\"o}tting}}, \bibinfo {author} {\bibfnamefont {F.}~\bibnamefont {Lohof}},\ and\ \bibinfo {author} {\bibfnamefont {C.}~\bibnamefont {Gies}},\ }\href {https://doi.org/10.1103/PhysRevB.105.165419} {\bibfield  {journal} {\bibinfo  {journal} {Phys. Rev. B}\ }\textbf {\bibinfo {volume} {105}},\ \bibinfo {pages} {165419} (\bibinfo {year} {2022})}\BibitemShut {NoStop}%
\bibitem [{\citenamefont {Marzari}\ and\ \citenamefont {Vanderbilt}(1997)}]{marzari1997}%
  \BibitemOpen
  \bibfield  {author} {\bibinfo {author} {\bibfnamefont {N.}~\bibnamefont {Marzari}}\ and\ \bibinfo {author} {\bibfnamefont {D.}~\bibnamefont {Vanderbilt}},\ }\href {https://doi.org/10.1103/PhysRevB.56.12847} {\bibfield  {journal} {\bibinfo  {journal} {Phys. Rev. B}\ }\textbf {\bibinfo {volume} {56}},\ \bibinfo {pages} {12847} (\bibinfo {year} {1997})}\BibitemShut {NoStop}%
\bibitem [{\citenamefont {Pizzi}\ \emph {et~al.}(2020)\citenamefont {Pizzi}, \citenamefont {Vitale}, \citenamefont {Arita}, \citenamefont {Blügel}, \citenamefont {Freimuth}, \citenamefont {Géranton}, \citenamefont {Gibertini}, \citenamefont {Gresch}, \citenamefont {Johnson}, \citenamefont {Koretsune}, \citenamefont {Ibañez-Azpiroz}, \citenamefont {Lee}, \citenamefont {Lihm}, \citenamefont {Marchand}, \citenamefont {Marrazzo}, \citenamefont {Mokrousov}, \citenamefont {Mustafa}, \citenamefont {Nohara}, \citenamefont {Nomura}, \citenamefont {Paulatto}, \citenamefont {Poncé}, \citenamefont {Ponweiser}, \citenamefont {Qiao}, \citenamefont {Thöle}, \citenamefont {Tsirkin}, \citenamefont {Wierzbowska}, \citenamefont {Marzari}, \citenamefont {Vanderbilt}, \citenamefont {Souza}, \citenamefont {Mostofi},\ and\ \citenamefont {Yates}}]{pizzi2020}%
  \BibitemOpen
  \bibfield  {author} {\bibinfo {author} {\bibfnamefont {G.}~\bibnamefont {Pizzi}}, \bibinfo {author} {\bibfnamefont {V.}~\bibnamefont {Vitale}}, \bibinfo {author} {\bibfnamefont {R.}~\bibnamefont {Arita}}, \bibinfo {author} {\bibfnamefont {S.}~\bibnamefont {Blügel}}, \bibinfo {author} {\bibfnamefont {F.}~\bibnamefont {Freimuth}}, \bibinfo {author} {\bibfnamefont {G.}~\bibnamefont {Géranton}}, \bibinfo {author} {\bibfnamefont {M.}~\bibnamefont {Gibertini}}, \bibinfo {author} {\bibfnamefont {D.}~\bibnamefont {Gresch}}, \bibinfo {author} {\bibfnamefont {C.}~\bibnamefont {Johnson}}, \bibinfo {author} {\bibfnamefont {T.}~\bibnamefont {Koretsune}}, \bibinfo {author} {\bibfnamefont {J.}~\bibnamefont {Ibañez-Azpiroz}}, \bibinfo {author} {\bibfnamefont {H.}~\bibnamefont {Lee}}, \bibinfo {author} {\bibfnamefont {J.-M.}\ \bibnamefont {Lihm}}, \bibinfo {author} {\bibfnamefont {D.}~\bibnamefont {Marchand}}, \bibinfo {author} {\bibfnamefont {A.}~\bibnamefont {Marrazzo}}, \bibinfo {author} {\bibfnamefont
  {Y.}~\bibnamefont {Mokrousov}}, \bibinfo {author} {\bibfnamefont {J.~I.}\ \bibnamefont {Mustafa}}, \bibinfo {author} {\bibfnamefont {Y.}~\bibnamefont {Nohara}}, \bibinfo {author} {\bibfnamefont {Y.}~\bibnamefont {Nomura}}, \bibinfo {author} {\bibfnamefont {L.}~\bibnamefont {Paulatto}}, \bibinfo {author} {\bibfnamefont {S.}~\bibnamefont {Poncé}}, \bibinfo {author} {\bibfnamefont {T.}~\bibnamefont {Ponweiser}}, \bibinfo {author} {\bibfnamefont {J.}~\bibnamefont {Qiao}}, \bibinfo {author} {\bibfnamefont {F.}~\bibnamefont {Thöle}}, \bibinfo {author} {\bibfnamefont {S.~S.}\ \bibnamefont {Tsirkin}}, \bibinfo {author} {\bibfnamefont {M.}~\bibnamefont {Wierzbowska}}, \bibinfo {author} {\bibfnamefont {N.}~\bibnamefont {Marzari}}, \bibinfo {author} {\bibfnamefont {D.}~\bibnamefont {Vanderbilt}}, \bibinfo {author} {\bibfnamefont {I.}~\bibnamefont {Souza}}, \bibinfo {author} {\bibfnamefont {A.~A.}\ \bibnamefont {Mostofi}},\ and\ \bibinfo {author} {\bibfnamefont {J.~R.}\ \bibnamefont {Yates}},\ }\href
  {https://doi.org/10.1088/1361-648X/ab51ff} {\bibfield  {journal} {\bibinfo  {journal} {Journal of Physics: Condensed Matter}\ }\textbf {\bibinfo {volume} {32}},\ \bibinfo {pages} {165902} (\bibinfo {year} {2020})}\BibitemShut {NoStop}%
\bibitem [{\citenamefont {Gerhard}\ \emph {et~al.}(2026)\citenamefont {Gerhard}, \citenamefont {Wang}, \citenamefont {Cerjan},\ and\ \citenamefont {Benalcazar}}]{gerhard2026}%
  \BibitemOpen
  \bibfield  {author} {\bibinfo {author} {\bibfnamefont {H.}~\bibnamefont {Gerhard}}, \bibinfo {author} {\bibfnamefont {Y.}~\bibnamefont {Wang}}, \bibinfo {author} {\bibfnamefont {A.}~\bibnamefont {Cerjan}},\ and\ \bibinfo {author} {\bibfnamefont {W.~A.}\ \bibnamefont {Benalcazar}},\ }\href {https://arxiv.org/abs/2603.13206} {\bibinfo {title} {A spatial localizer for electrons in insulators}} (\bibinfo {year} {2026}),\ \Eprint {https://arxiv.org/abs/2603.13206} {arXiv:2603.13206 [cond-mat.mtrl-sci]} \BibitemShut {NoStop}%
\bibitem [{\citenamefont {Loring}(2015)}]{loring2015}%
  \BibitemOpen
  \bibfield  {author} {\bibinfo {author} {\bibfnamefont {T.~A.}\ \bibnamefont {Loring}},\ }\bibfield  {journal} {\bibinfo  {journal} {Annals of Physics}\ }\textbf {\bibinfo {volume} {356}},\ \href {https://doi.org/10.1016/j.aop.2015.02.031} {10.1016/j.aop.2015.02.031} (\bibinfo {year} {2015})\BibitemShut {NoStop}%
\bibitem [{\citenamefont {Cerjan}\ and\ \citenamefont {Loring}(2022)}]{cerjan2022}%
  \BibitemOpen
  \bibfield  {author} {\bibinfo {author} {\bibfnamefont {A.}~\bibnamefont {Cerjan}}\ and\ \bibinfo {author} {\bibfnamefont {T.~A.}\ \bibnamefont {Loring}},\ }\href {https://doi.org/10.1103/PhysRevB.106.064109} {\bibfield  {journal} {\bibinfo  {journal} {Phys. Rev. B}\ }\textbf {\bibinfo {volume} {106}},\ \bibinfo {pages} {064109} (\bibinfo {year} {2022})}\BibitemShut {NoStop}%
\bibitem [{\citenamefont {Jankowski}\ \emph {et~al.}(2025)\citenamefont {Jankowski}, \citenamefont {Thompson}, \citenamefont {Monserrat},\ and\ \citenamefont {Slager}}]{jankowski2025}%
  \BibitemOpen
  \bibfield  {author} {\bibinfo {author} {\bibfnamefont {W.~J.}\ \bibnamefont {Jankowski}}, \bibinfo {author} {\bibfnamefont {J.~J.~P.}\ \bibnamefont {Thompson}}, \bibinfo {author} {\bibfnamefont {B.}~\bibnamefont {Monserrat}},\ and\ \bibinfo {author} {\bibfnamefont {R.-J.}\ \bibnamefont {Slager}},\ }\href {https://doi.org/10.1038/s41467-025-59257-5} {\bibfield  {journal} {\bibinfo  {journal} {Nat. Commun.}\ }\textbf {\bibinfo {volume} {16}},\ \bibinfo {pages} {4661} (\bibinfo {year} {2025})}\BibitemShut {NoStop}%
\bibitem [{\citenamefont {Thompson}\ \emph {et~al.}(2025)\citenamefont {Thompson}, \citenamefont {Jankowski}, \citenamefont {Slager},\ and\ \citenamefont {Monserrat}}]{thompson2025}%
  \BibitemOpen
  \bibfield  {author} {\bibinfo {author} {\bibfnamefont {J.~J.~P.}\ \bibnamefont {Thompson}}, \bibinfo {author} {\bibfnamefont {W.~J.}\ \bibnamefont {Jankowski}}, \bibinfo {author} {\bibfnamefont {R.-J.}\ \bibnamefont {Slager}},\ and\ \bibinfo {author} {\bibfnamefont {B.}~\bibnamefont {Monserrat}},\ }\href {https://doi.org/10.1038/s41467-025-66276-9} {\bibfield  {journal} {\bibinfo  {journal} {Nat. Commun.}\ }\textbf {\bibinfo {volume} {16}},\ \bibinfo {pages} {11448} (\bibinfo {year} {2025})}\BibitemShut {NoStop}%
\bibitem [{\citenamefont {Paiva}\ \emph {et~al.}(2024)\citenamefont {Paiva}, \citenamefont {Holder},\ and\ \citenamefont {Ilan}}]{paiva2024}%
  \BibitemOpen
  \bibfield  {author} {\bibinfo {author} {\bibfnamefont {C.}~\bibnamefont {Paiva}}, \bibinfo {author} {\bibfnamefont {T.}~\bibnamefont {Holder}},\ and\ \bibinfo {author} {\bibfnamefont {R.}~\bibnamefont {Ilan}},\ }\href@noop {} {\bibinfo {title} {Shift and polarization of excitons from quantum geometry}} (\bibinfo {year} {2024}),\ \Eprint {https://arxiv.org/abs/2408.10300} {arXiv:2408.10300 [cond-mat.mes-hall]} \BibitemShut {NoStop}%
\bibitem [{\citenamefont {Cao}\ \emph {et~al.}(2021)\citenamefont {Cao}, \citenamefont {Fertig},\ and\ \citenamefont {Brey}}]{cao2021}%
  \BibitemOpen
  \bibfield  {author} {\bibinfo {author} {\bibfnamefont {J.}~\bibnamefont {Cao}}, \bibinfo {author} {\bibfnamefont {H.~A.}\ \bibnamefont {Fertig}},\ and\ \bibinfo {author} {\bibfnamefont {L.}~\bibnamefont {Brey}},\ }\href {https://doi.org/10.1103/PhysRevB.103.115422} {\bibfield  {journal} {\bibinfo  {journal} {Phys. Rev. B}\ }\textbf {\bibinfo {volume} {103}},\ \bibinfo {pages} {115422} (\bibinfo {year} {2021})}\BibitemShut {NoStop}%
\bibitem [{\citenamefont {Fertig}\ and\ \citenamefont {Brey}(2025)}]{fertig2025}%
  \BibitemOpen
  \bibfield  {author} {\bibinfo {author} {\bibfnamefont {H.~A.}\ \bibnamefont {Fertig}}\ and\ \bibinfo {author} {\bibfnamefont {L.}~\bibnamefont {Brey}},\ }\href {https://doi.org/10.1103/PhysRevB.111.035158} {\bibfield  {journal} {\bibinfo  {journal} {Phys. Rev. B}\ }\textbf {\bibinfo {volume} {111}},\ \bibinfo {pages} {035158} (\bibinfo {year} {2025})}\BibitemShut {NoStop}%
\bibitem [{\citenamefont {Davenport}\ \emph {et~al.}(2026{\natexlab{b}})\citenamefont {Davenport}, \citenamefont {Hwang}, \citenamefont {Knolle},\ and\ \citenamefont {Schindler}}]{davenport2026composite}%
  \BibitemOpen
  \bibfield  {author} {\bibinfo {author} {\bibfnamefont {H.}~\bibnamefont {Davenport}}, \bibinfo {author} {\bibfnamefont {Y.}~\bibnamefont {Hwang}}, \bibinfo {author} {\bibfnamefont {J.}~\bibnamefont {Knolle}},\ and\ \bibinfo {author} {\bibfnamefont {F.}~\bibnamefont {Schindler}},\ }\href@noop {} {\bibinfo {title} {Composite quantum geometry and semiclassical dynamics}} (\bibinfo {year} {2026}{\natexlab{b}}),\ \Eprint {https://arxiv.org/abs/2606.12525} {arXiv:2606.12525 [cond-mat.mes-hall]} \BibitemShut {NoStop}%
\bibitem [{\citenamefont {Mendez}\ \emph {et~al.}(2026)\citenamefont {Mendez}, \citenamefont {Brey},\ and\ \citenamefont {Fertig}}]{mendez2026}%
  \BibitemOpen
  \bibfield  {author} {\bibinfo {author} {\bibfnamefont {F.~I.}\ \bibnamefont {Mendez}}, \bibinfo {author} {\bibfnamefont {L.}~\bibnamefont {Brey}},\ and\ \bibinfo {author} {\bibfnamefont {H.~A.}\ \bibnamefont {Fertig}},\ }\href {https://arxiv.org/abs/2605.22810} {\bibinfo {title} {Signatures of the quantum geometric dipole of interlayer excitons in counterflow conductivity}} (\bibinfo {year} {2026}),\ \Eprint {https://arxiv.org/abs/2605.22810} {arXiv:2605.22810 [cond-mat.mes-hall]} \BibitemShut {NoStop}%
\bibitem [{\citenamefont {Chen}\ \emph {et~al.}(2026)\citenamefont {Chen}, \citenamefont {Ghorashi}, \citenamefont {Cano},\ and\ \citenamefont {Crépel}}]{chen2026}%
  \BibitemOpen
  \bibfield  {author} {\bibinfo {author} {\bibfnamefont {L.}~\bibnamefont {Chen}}, \bibinfo {author} {\bibfnamefont {S.~A.~A.}\ \bibnamefont {Ghorashi}}, \bibinfo {author} {\bibfnamefont {J.}~\bibnamefont {Cano}},\ and\ \bibinfo {author} {\bibfnamefont {V.}~\bibnamefont {Crépel}},\ }\href {https://arxiv.org/abs/2506.22417} {\bibinfo {title} {Quantum-geometric dipole: a topological boost to flavor ferromagnetism in flat bands}} (\bibinfo {year} {2026}),\ \Eprint {https://arxiv.org/abs/2506.22417} {arXiv:2506.22417 [cond-mat.mes-hall]} \BibitemShut {NoStop}%
\bibitem [{\citenamefont {Yang}\ \emph {et~al.}(2026{\natexlab{a}})\citenamefont {Yang}, \citenamefont {Zheng}, \citenamefont {Xu}, \citenamefont {Xiao},\ and\ \citenamefont {Cao}}]{yang2026gianthelicaldipole}%
  \BibitemOpen
  \bibfield  {author} {\bibinfo {author} {\bibfnamefont {K.}~\bibnamefont {Yang}}, \bibinfo {author} {\bibfnamefont {H.}~\bibnamefont {Zheng}}, \bibinfo {author} {\bibfnamefont {X.}~\bibnamefont {Xu}}, \bibinfo {author} {\bibfnamefont {D.}~\bibnamefont {Xiao}},\ and\ \bibinfo {author} {\bibfnamefont {T.}~\bibnamefont {Cao}},\ }\href {https://arxiv.org/abs/2604.12295} {\bibinfo {title} {Giant and helical exciton dipole from berry curvature in flat chern bands}} (\bibinfo {year} {2026}{\natexlab{a}}),\ \Eprint {https://arxiv.org/abs/2604.12295} {arXiv:2604.12295 [cond-mat.mes-hall]} \BibitemShut {NoStop}%
\bibitem [{\citenamefont {Yang}\ \emph {et~al.}(2026{\natexlab{b}})\citenamefont {Yang}, \citenamefont {Srivastava},\ and\ \citenamefont {Song}}]{yang2026shift}%
  \BibitemOpen
  \bibfield  {author} {\bibinfo {author} {\bibfnamefont {X.}~\bibnamefont {Yang}}, \bibinfo {author} {\bibfnamefont {A.}~\bibnamefont {Srivastava}},\ and\ \bibinfo {author} {\bibfnamefont {J.~C.~W.}\ \bibnamefont {Song}},\ }\href {https://doi.org/10.1038/s41467-026-72878-8} {\bibfield  {journal} {\bibinfo  {journal} {Nat. Commun.}\ }\textbf {\bibinfo {volume} {17}},\ \bibinfo {pages} {6465} (\bibinfo {year} {2026}{\natexlab{b}})}\BibitemShut {NoStop}%
\bibitem [{\citenamefont {Hu}\ \emph {et~al.}(2026)\citenamefont {Hu}, \citenamefont {Kundu}, \citenamefont {Guo}, \citenamefont {Thompson}, \citenamefont {Li}, \citenamefont {Wang},\ and\ \citenamefont {Monserrat}}]{hu2026shift}%
  \BibitemOpen
  \bibfield  {author} {\bibinfo {author} {\bibfnamefont {J.}~\bibnamefont {Hu}}, \bibinfo {author} {\bibfnamefont {S.}~\bibnamefont {Kundu}}, \bibinfo {author} {\bibfnamefont {Z.}~\bibnamefont {Guo}}, \bibinfo {author} {\bibfnamefont {J.~J.~P.}\ \bibnamefont {Thompson}}, \bibinfo {author} {\bibfnamefont {W.}~\bibnamefont {Li}}, \bibinfo {author} {\bibfnamefont {H.}~\bibnamefont {Wang}},\ and\ \bibinfo {author} {\bibfnamefont {B.}~\bibnamefont {Monserrat}},\ }\href@noop {} {\bibinfo {title} {Generalized shift vector as the intrinsic dipole of many-body correlated electronic states}} (\bibinfo {year} {2026}),\ \Eprint {https://arxiv.org/abs/2605.23431} {arXiv:2605.23431 [cond-mat.mes-hall]} \BibitemShut {NoStop}%
\bibitem [{\citenamefont {Maccone}\ and\ \citenamefont {Pati}(2014)}]{maccone_uncertainty}%
  \BibitemOpen
  \bibfield  {author} {\bibinfo {author} {\bibfnamefont {L.}~\bibnamefont {Maccone}}\ and\ \bibinfo {author} {\bibfnamefont {A.~K.}\ \bibnamefont {Pati}},\ }\href {https://doi.org/10.1103/PhysRevLett.113.260401} {\bibfield  {journal} {\bibinfo  {journal} {Phys. Rev. Lett.}\ }\textbf {\bibinfo {volume} {113}},\ \bibinfo {pages} {260401} (\bibinfo {year} {2014})}\BibitemShut {NoStop}%
\bibitem [{\citenamefont {Alexandradinata}\ \emph {et~al.}(2014)\citenamefont {Alexandradinata}, \citenamefont {Dai},\ and\ \citenamefont {Bernevig}}]{alexandradinata2014}%
  \BibitemOpen
  \bibfield  {author} {\bibinfo {author} {\bibfnamefont {A.}~\bibnamefont {Alexandradinata}}, \bibinfo {author} {\bibfnamefont {X.}~\bibnamefont {Dai}},\ and\ \bibinfo {author} {\bibfnamefont {B.~A.}\ \bibnamefont {Bernevig}},\ }\href {https://doi.org/10.1103/PhysRevB.89.155114} {\bibfield  {journal} {\bibinfo  {journal} {Phys. Rev. B}\ }\textbf {\bibinfo {volume} {89}},\ \bibinfo {pages} {155114} (\bibinfo {year} {2014})}\BibitemShut {NoStop}%
\bibitem [{\citenamefont {Tao}\ \emph {et~al.}(2026)\citenamefont {Tao}, \citenamefont {Haber},\ and\ \citenamefont {Neaton}}]{tao2025wfdx}%
  \BibitemOpen
  \bibfield  {author} {\bibinfo {author} {\bibfnamefont {Z.}~\bibnamefont {Tao}}, \bibinfo {author} {\bibfnamefont {J.~B.}\ \bibnamefont {Haber}},\ and\ \bibinfo {author} {\bibfnamefont {J.~B.}\ \bibnamefont {Neaton}},\ }\href {https://doi.org/10.1021/acs.jctc.5c01686} {\bibfield  {journal} {\bibinfo  {journal} {Journal of Chemical Theory and Computation}\ }\textbf {\bibinfo {volume} {22}},\ \bibinfo {pages} {588} (\bibinfo {year} {2026})}\BibitemShut {NoStop}%
\bibitem [{\citenamefont {Resta}(1998)}]{resta1998}%
  \BibitemOpen
  \bibfield  {author} {\bibinfo {author} {\bibfnamefont {R.}~\bibnamefont {Resta}},\ }\href {https://doi.org/10.1103/PhysRevLett.80.1800} {\bibfield  {journal} {\bibinfo  {journal} {Phys. Rev. Lett.}\ }\textbf {\bibinfo {volume} {80}},\ \bibinfo {pages} {1800} (\bibinfo {year} {1998})}\BibitemShut {NoStop}%
\bibitem [{\citenamefont {Rohlfing}\ and\ \citenamefont {Louie}(2000)}]{rohlfing2000}%
  \BibitemOpen
  \bibfield  {author} {\bibinfo {author} {\bibfnamefont {M.}~\bibnamefont {Rohlfing}}\ and\ \bibinfo {author} {\bibfnamefont {S.~G.}\ \bibnamefont {Louie}},\ }\href {https://doi.org/10.1103/PhysRevB.62.4927} {\bibfield  {journal} {\bibinfo  {journal} {Phys. Rev. B}\ }\textbf {\bibinfo {volume} {62}},\ \bibinfo {pages} {4927} (\bibinfo {year} {2000})}\BibitemShut {NoStop}%
\bibitem [{\citenamefont {Brauer}\ and\ \citenamefont {Weyl}(1935)}]{brauer1935}%
  \BibitemOpen
  \bibfield  {author} {\bibinfo {author} {\bibfnamefont {R.}~\bibnamefont {Brauer}}\ and\ \bibinfo {author} {\bibfnamefont {H.}~\bibnamefont {Weyl}},\ }\href {http://www.jstor.org/stable/2371218} {\bibfield  {journal} {\bibinfo  {journal} {American Journal of Mathematics}\ }\textbf {\bibinfo {volume} {57}},\ \bibinfo {pages} {425} (\bibinfo {year} {1935})}\BibitemShut {NoStop}%
\bibitem [{\citenamefont {Chiu}\ \emph {et~al.}(2016)\citenamefont {Chiu}, \citenamefont {Teo}, \citenamefont {Schnyder},\ and\ \citenamefont {Ryu}}]{chiu2016}%
  \BibitemOpen
  \bibfield  {author} {\bibinfo {author} {\bibfnamefont {C.-K.}\ \bibnamefont {Chiu}}, \bibinfo {author} {\bibfnamefont {J.~C.~Y.}\ \bibnamefont {Teo}}, \bibinfo {author} {\bibfnamefont {A.~P.}\ \bibnamefont {Schnyder}},\ and\ \bibinfo {author} {\bibfnamefont {S.}~\bibnamefont {Ryu}},\ }\href {https://doi.org/10.1103/RevModPhys.88.035005} {\bibfield  {journal} {\bibinfo  {journal} {Rev. Mod. Phys.}\ }\textbf {\bibinfo {volume} {88}},\ \bibinfo {pages} {035005} (\bibinfo {year} {2016})}\BibitemShut {NoStop}%
\bibitem [{\citenamefont {Zirnbauer}(2021)}]{zirnbauer2021}%
  \BibitemOpen
  \bibfield  {author} {\bibinfo {author} {\bibfnamefont {M.~R.}\ \bibnamefont {Zirnbauer}},\ }\href {https://doi.org/10.1063/5.0035358} {\bibfield  {journal} {\bibinfo  {journal} {Journal of Mathematical Physics}\ }\textbf {\bibinfo {volume} {62}},\ \bibinfo {pages} {021101} (\bibinfo {year} {2021})}\BibitemShut {NoStop}%
\end{thebibliography}%

\end{document}